\documentclass{aa}  

\usepackage{graphicx}
\usepackage{txfonts}
\usepackage{enumitem}
\usepackage{caption}
\usepackage{subcaption}
\usepackage{float}
\usepackage{fancyvrb}
\usepackage{hyperref}
\hypersetup{colorlinks=true,linkcolor=blue,citecolor=blue,urlcolor=blue}
\usepackage{amsmath}

\usepackage{stfloats}

\begin{document} 

   \title{Algorithms and radiation dynamics for the vicinity of black holes\\
   II. Results}

   \author{Leela Elpida Koutsantoniou}

   \institute{Department of Astrophysics, Astronomy and Mechanics, Faculty of Physics, University of Athens, Panepistimiopolis Zografos, Athens 15784, Greece\newline
   e-mail: \texttt{leelamk@phys.uoa.gr}}
    \date{Received 21 June 2022 / Accepted 18 January 2023}

\abstract{
We present the results of our studies on accretion disks in the proximity of astrophysical black holes. These disks can be of varying degrees of opacity, geometrical shapes, sizes, and volumes. The central compact object is a Schwarzschild or a Kerr black hole of various spin parameters. We describe the environment and the physics of the systems under examination and the disk models considered. We first investigate the effects of the spacetime rotation on photon trajectories. We then examine the radiation forces recorded at various points of the arrangement inside and outside the disk material, and in the inner, outer, and off-equatorial material orbits. We document and explore the radiation effects, which are revealed to be significant and positively consequential. Afterward, we inspect the possible imaging outcome of various types of black hole and accretion disk configurations, and we show our results for plots that could be used to estimate the central black hole spin in a system. Finally, we show results regarding the disk material orbit degradation due to its thermal radiation.
}


   \maketitle

%-------------------------------------------------------------------

\section{Introduction}
Black hole (BH) observations and research have made considerable advances in recent years. Making use of technological developments, new ground instruments {are being constructed and updated,} and state-of-the-art space programs designed to observe and record these perplexing objects and their peculiar environment. Even instruments not intended for BH observation are often used for such a purpose since they can document the effects of secondary phenomena originating from a compact object and its system. These instruments can thus frequently provide information about the object under study and its properties.

The mass, $M$, of a stellar-mass BH is estimated to range between approximately 5 and 100 {Solar} masses (e.g., \citealt{stellarbh}\footnote{The lower limit for BH mass, the Tolman–Oppenheimer–Volkoff limit is still under active research, see \citet{bombaci,kaloger-baum,thompson19,abbott}, {but also \citet{unicorn} and \citet{giraffe}.}}).
In most cases, we observe stellar-mass BHs surrounded by accretion disks (ADs) for varying periods of times, usually from a few days to a few hundred days, depending on the configuration setup and the greatly varying accreting mass quantity (see, e.g., \citealp{Lasota16}). We thus turned our attention to such systems and investigated the assorted radiation effects encountered in strong gravity environments, examining their properties and the radiation impact.

The temperature of ADs of this type is fairly high (${T~\rm \sim10^7}$K) and causes the emission of X-rays at $E~\rm \sim1keV$ or higher \citep{Zhang,KCKC}. The presence of these noncentral sources of light brings into play dynamical effects that modify the equilibrium and shape the evolution of the disk. The primary goal of our research was to study the Poynting--Robertson (PR) effect \citep{Poynting,Larmor,Page1918b,Robertson} in BH and AD systems. We then evaluated the ramifications of the PR effect in these setups and assessed the importance of the induced effects. This phenomenon is noteworthy because it creates a drag force that acts on the disk material. Thereupon, the material slowly brakes, falls farther inward, and finally accretes onto the central compact object. The nature of strong gravity environments{, however,} sets in motion a series of more complicated and entangled effects that require proper physical treatment.

In the first part of this work \citep{KpartI}, we explained the theoretical formulation and the equations required to study the problem. We also described the AD models we used in that study, presented the codes written and used for this research, and gave an account of their capabilities. In this second part, we show and explain the results of the various facets of our study and discuss their implications. We begin by investigating the ramifications of the {spacetime} curvature and its effects on photon trajectories. We then examine the recorded radiation forces on the AD material for various models and spin parameters. We should keep in mind here that these radiation forces depend on the target velocity and hence have distinct consequences on particles of different velocity profiles. We thus consider the significance and the importance of the disk material characteristics. Further on, we investigate our BH-AD imaging capabilities and examine what additional information can be drawn from this for the systems. Finally, we discern the subsequent effects of the radiation pressure on massive particle trajectories.

Some of the first and perhaps most fundamental steps in the study of radiation in the vicinity of massive objects were taken by \citet*{AEL90}, \citet{ML93,ML96}, and \citet{LM95}. These groups studied the radiation of a central star that reaches specific points in the environment and presented analytical formulae that calculate the radiation field in these potentially less complicated conditions. We should also mention the noteworthy studies presented in \citet{FW04,FW07} and \citet{WF08a,WF08b}, where radiation, radiative transfer, and emission lines in proximity to BHs were surveyed assiduously. The relevant work presented in \citet{YWF} and \citet{YounsiPhD}, which extended the aforementioned studies, is also worth mentioning. Furthermore, thorough research on the numerous and intricate aspects of radiation in strong gravity was conducted by \citet{Bini09,BdFG,Bini11,Bini15}, who examined a notable number of intriguing subjects.

Our current study was initiated to better examine the {'cosmic battery' mechanism} \citep{CB98,CKC06,ic09} and evaluate its effects on ADs around BHs. In this process, the hot disk orbiting the BH radiates thermally due to its high temperature ($T\rm \gtrsim10^7K, E\rm \gtrsim1keV$; see, e.g., \citealt{KCKC}). This radiation is then absorbed by the disk material itself, exerting radiation forces influenced by the aforementioned PR effect. The radiation{, nonetheless,} primarily affects the material electrons since ${f_{e}/f_{p}\sim(m_{p}/m_{e})^{2}}$, where ${{f}_{p}}$ and ${{f}_{e}}$ are the force on a proton and on an electron and ${{m}_{p}}$ and ${{m}_{e}}$ are their masses, respectively. The consequence of this is the electrons moving with a lower speed than the protons and thus the generation of a ring current. This ring current in turn generates a poloidal magnetic field in the system. The presence of a growing and expanding magnetic field in this environment has further effects and implications. These include, {among others,} the Blandford--Znajek process \citep{BZ,Livio99,Komissarov2001,McKinney2005}, which induces the extraction of energy from the BH. All these new physical components and circumstances then lead to different structural forms and equilibrium conditions in the entire configuration, which should be investigated anew.

The best environments where such situations and phenomena can be examined are systems of binary stars that include one typical “donor” star (a main sequence star, red giant, or white dwarf) and one “accretor” compact star (a neutron star or BH). These X-ray binary systems are illuminated due to matter transference from the more gas-rich companion to the compact component. During this process, matter is stripped off the companion star’s external layers and creates a short-lived AD around the compact object (e.g., \citealp{Lasota16}). This reconfiguration compels the infalling matter to release its gravitational potential energy and emit it as luminous energy in the X-ray band. Abundant theoretical and observational information concerning these intriguing arrangements can be found, for instance, in \citet{Pods02}, \citet{Tauris06}, \citet{BellBook10}, \citet{Bell20B}, \citet{Bell10A}, \citet{KCKC}, and \citet{Tet}. As for the X-ray binaries' evolutionary sequence, periodicity, and variability timescales, we can find detailed information for the various different situations in \citet{vdk89}, \citet{vdk95}, \citet{vdk04}, \citet{Uttley}, \citet{Koljonen}, \citet{reigA}, \citet{reigB}, and \citet{Lasota16}.

We should also mention some of the notable observational studies that provided valuable information for our research and numerous other related studies. These include observations made using
{Very-long-baseline interferometry} (VLBI; e.g., \citealp{zt03,tz10}), the Very Long Baseline Array (VLBA; e.g., \citealp{zt02,zt03,zt04,zt05,tz10}), the Monitoring Of Jets in Active galactic nuclei with VLBA Experiments (MOJAVE) program (e.g., \citealt{mojave8,mojave11}), the \textit{Hubble} Space Telescope (HST; e.g., \citealt{Tsvetanov}), the Very Large Array (VLA; e.g., \citealt{zt02,zt04,Herrnstein, carilli,kl04,Govoni}), \textit{Chandra} (e.g., \citealp{Kronberg} and references therein), the INTErnational Gamma-Ray Astrophysics Laboratory (INTEGRAL) and \textit{Swift} (e.g., \citealp{Bonafede} and references therein), the Röntgensatellit (ROSAT; e.g., \citealp{widrow}), the X-ray Multi-Mirror Mission (\textit{XMM}-\textit{Newton}; e.g., \citealp{widrow} and references therein), Skinakas Observatory (SKO; e.g., \citealp{reigA,reigB}), %MAXI
and the Event Horizon Telescope (EHT; e.g., M87: \citealp{EHTa1,EHTa4,EHTa6,EHTa7,EHTa8}, Sgr A*: \citealp{EHTb1,EHTb2,EHTb3,EHTb4}). Finally, and importantly, we should mention studies that combine a multitude of the aforementioned observatories and instruments in order to examine phenomena closely related to, and very useful for, our research. {These} include \citet{G04,G08,G12,G14,G15B,G15A}, \citet{G06}, \citet{Gbook}, \citet{MG07,MG08,MG09}, \citet{MahmudGB09}, \citet{ic09}, \citet{OG09}, \citet{RG12}, \citet{Murphy13A}, \citet{Mahmud13}, \citet{Murphy13B}, and \citet{Christodoulou16}.

In the present work, we start in Sect. \ref{setting up} by setting up the problem: we show the Kerr metric and its notable surfaces, the general relativistic radiative transfer equation (GRRTE) and its solution, and the AD models we used. In Sect. \ref{Results} we present the results of our studies: photon trajectories in strong gravity, radiation forces and dynamics, BH imaging at infinity and its capabilities, and massive particle trajectories in strong gravity under the influence of radiation pressure. Finally, in Sect. \ref{Conclusions} we summarize our results and discuss their implications.

\section{Setting up the problem}
\label{setting up}

\subsection{The Kerr metric and notable surfaces}
\label{Kerr notable}

We assumed that the immediate environment around a potentially rotating and accreting BH can be adequately described using the Kerr metric \citep{Kerr,B72}. The setup's {spacetime} can thus be determined by the central compact object, {which} is axisymmetric, uncharged, and possibly rotating. Additionally, we assumed that the presence and motion of test particles does not affect the {spacetime} form or the stress--energy tensor. We used the geometrized unit system, where ${c=G=1}$ and distances are measured in units of gravitational radii, $r_{g}=G M/c^{2}=M$, where $M$ is the BH mass. We also assumed the Einstein notation for summation over double indices. We denote {spacetime} components with Greek indices and space components with Latin indices.

We can fully describe the BH and the {spacetime} around it using its mass, $M$, and spin parameter, $a$. The spin parameter, $a$, takes values from $a=0$ for a nonrotating Schwarzschild BH to $a=M$ for a maximally rotating one. Alternatively, we can use $j=0$ and $j=1,$ respectively, for the dimensionless angular momentum $j=a/M$. The Kerr metric in Boyer–Lindquist (BL) coordinates $(t,\phi,r,\theta)$ is then given by\begin{eqnarray}
{\rm d}{{s}^{2}}&=&
{{g}_{\alpha \beta }} {\rm d}{{x}^{\alpha }}{\rm d}{{x}^{\beta }} \ \nonumber\\
&=&-{{e}^{2\nu }}{\rm d}{{t}^{2}}+{{e}^{2\psi }}{{\left( {\rm d}\phi -\omega {\rm d}t \right)}^{2}}+{{e}^{2{{\mu }_{1}}}}{\rm d}{{r}^{2}}+{{e}^{2{{\mu }_{2}}}}{\rm d}{{\theta }^{2}}\ , \label{metric}
\end{eqnarray}
\noindent where
\begin{equation}
{{e}^{2\nu }}=\frac{\Sigma \Delta }{A}\ ,\
{{e}^{2\psi }}=\frac{A{{\sin}^{2}}\theta }{\Sigma }\ ,\
{{e}^{2{{\mu }_{1}}}}=\frac{\Sigma }{\Delta }\ ,\
{{e}^{2{{\mu }_{2}}}}=\Sigma\ ,
\end{equation}
\newpage
\noindent with

\begin{equation}
    \begin{aligned}
    \Delta &={{r}^{2}}-2Mr+{{a}^{2}}\ ,\\
    \Sigma &={{r}^{2}}+{{a}^{2}}{{\cos}^{2}}\theta\ ,\\
    A &={{\left( {{r}^{2}}+{{a}^{2}} \right)}^{2}}-{{a}^{2}}\Delta {{\sin}^{2}}\theta\ ,
    \end{aligned}
\end{equation}

\noindent and the {spacetime} angular velocity is given by
\begin{equation}
    \omega =-\frac{{{g}_{\phi t}}}{{{g}_{\phi \phi}}}=\frac{2Mra}{A}\ \
    \label{omega spacetime}
\end{equation}
(see \citealt{B70} and \citealt*{B72}).

% evh_ergo3.png --------------------------------------------
   \begin{figure}
   \centering
   \includegraphics[width=\hsize]{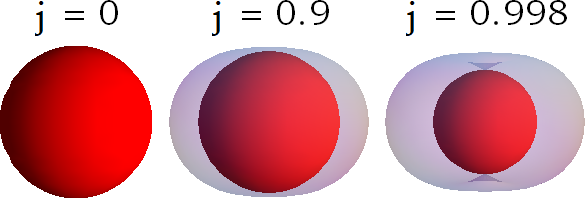}
      \caption{Event horizon (in red) and ergosphere (in gray) of a BH with various angular momentum values. The ergosphere starts in low $j$ as an ellipsoid, turning into a spindle torus for high $j$. We also notice the shrinkage of the event horizon for increasing values of $j$.
      }
      \label{evh_ergo}
   \end{figure}

From the metric (Eq. \ref{metric}) we {determine} assorted notable surfaces around a rotating BH. The BH outer event horizon firstly stems from one of the poles of the $g_{rr}$ component and can be found at the outermost root of the equation $\Delta = 0$:
\begin{equation}
    r_{\rm evh} = M+\sqrt{M^{2}-a^{2}}.
\end{equation}
The event horizon hence has the shape of a sphere with radius $r_{\rm evh} = 2M$ for a nonrotating Schwarzschild BH and with ${r_{\rm evh} = M}$ for a maximally rotating one.

The next notable surface is the static limit that establishes the outer boundary surface of the ergosphere. The static limit is found at the point where the $g_{tt}$ metric component changes sign, namely for radius
\begin{equation}
    {{r}_{\rm ergo}}=M+\sqrt{{{M}^{2}}-{{a}^{2}}{{\cos }^{2}}\theta }.
\end{equation}
For a Schwarzschild BH, the ergosphere is a sphere with radius ${r}_{\rm ergo}=2M$, the Schwarzschild radius. For increasing values of $a$, the ergosphere turns into an oblate spheroid with an equatorial (maximum) radius $r_{\rm ergo}^{\rm \theta=\pi/2}=2M$ and a polar (minimum) radius $r_{\rm ergo}^{\rm polar}=r_{\rm evh}$. Lastly, for a maximal BH, the ergosphere approximates a spindle torus\footnote{A spindle torus is a three-dimensional geometrical shape that is produced by the rotation of a circle around an axis in the same plane as the circle. The distance of this axis from the center of the circle is smaller than the circle radius. Objects that approximate a spindle torus are apples and pumpkins.} (Fig. \ref{evh_ergo}). From the above, we deduce that the ergosphere may change its shape for increasing $a$, yet it maintains a constant equatorial radius and a decreasing radius in the poles, where it always remains in contact with the event horizon. Inside this static limit, we have $g_{tt}<0$. This forces any particle moving there, massive or massless, to corotate with the BH with at least $\omega=-g_{\phi t}/g_{\phi \phi}$, so its proper time, $\tau$, remains positive. Finally, we also note that in theory it is possible to extract energy and mass from this region around the BH via the Penrose process \citep{Penrose71}.
   
Another relevant surface around the BH is the innermost circular equatorial orbit for particles. The limiting case is the one describing particle motion with infinite energy per unit mass. This is the photon orbit, the innermost equatorial circular orbit for any particle. The radius of the photon orbit is given by
\begin{equation}
    r_{\rm ph}=2M \Bigg\{1+\cos \left[\frac{2}{3}\: \cos^{-1}\left(\mp \frac{a}{M}\right)\right]\Bigg\},
\end{equation}
where the upper sign refers to direct orbits (corotating with the BH) and the lower sign to retrograde (counter-rotating) orbits. For a Schwarzschild BH ($a=0$), the photon orbit radius is $r_{\rm ph}=3M$; for a maximally rotating BH ($a=M$), the radii are $r_{\rm ph}=M$ for the direct and $r_{\rm ph}^{\rm retro}=4M$ for the retrograde photon ring.

Finally, we note the innermost stable circular orbit (ISCO) for massive particles of the disk. The radius of the ISCO is given by
\begin{equation}
   {{r}_{\rm ISCO}}=M\left[ 3+{{Z}_{2}}\mp \sqrt{\left( 3-{{Z}_{1}} \right)\left( 3+{{Z}_{1}}+2{{Z}_{2}} \right)} \right],
   \label{isco}
\end{equation}
where
\[
{{Z}_{1}}=1+{{\left( 1-\frac{{{a}^{2}}}{{{M}^{2}}} \right)}^{{1}/{3}\;}}\left[ {{\left( 1+\frac{a}{M} \right)}^{{1}/{3}\;}}+{{\left( 1-\frac{a}{M} \right)}^{{1}/{3}\;}} \right],
\]
\begin{equation}
    {{Z}_{2}}=\sqrt{\frac{3{{a}^{2}}}{{{M}^{2}}}+Z_{1}^{2}},
\end{equation}
where, again, the upper sign in Eq. (\ref{isco}) refers to direct orbits and the lower sign to retrograde orbits. The ISCO radius is hence ${r_{\rm ISCO}=6M}$ for $a=0$. For $a=M$, the direct ISCO has radius $r_{\rm ISCO}=M$, and the retrograde ISCO has radius $r_{\rm ISCO}^{\rm retro}=9M$.

\begin{figure}
   \centering
   \includegraphics[width=\hsize]{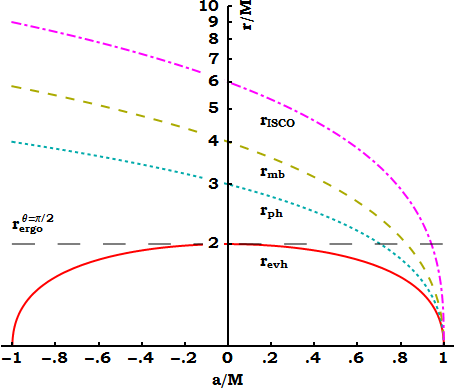}
      \caption{Radii of the notable equatorial orbits. In addition to the surfaces described in Sect. \ref{Kerr notable}, we also include for completeness the marginally bound circular orbit radius, $r_{\rm mb}$.
      }
      \label{orbit_plot}
   \end{figure}

The values and evolution of the radius of these notable surfaces for all spin parameters can be seen in Fig. \ref{orbit_plot}. For $a/M=0,$ we have a spherically symmetric object for all the surfaces. For $a/M>0$, we see the radius of the direct orbits and for $a/M<0$, the radius of the retrograde orbits. We should also note here that, even though the aforementioned characteristic surfaces appear to coincide at radius $r=M$ for a maximally rotating BH, in reality they do not. This is a false appearance, caused by the structure of the BL coordinate system. These metric solutions are distinct for any spin, $a$, and differ in radial proper distance \citep[e.g.,][]{B72,Chandra}.

Along with these notable orbits and surfaces, there are also other important structures. These include, for example, the marginally bound massive particle orbit with radius ${r_{\rm mb}=2M \mp a+2\sqrt{M(M \mp a)}}$ and the inner horizon or ergosphere. We do not use any of these, but more information on them can be found in \citet*{B72} or \citet*{MTW}.

We should also mention the coordinate angular velocity for a circular equatorial orbit, which is given by
\begin{equation}
    \Omega =\frac{{\rm d}\phi }{{\rm d}t}=\frac{{{u}^{\phi }}}{{{u}^{t}}}=\frac{{{M}^{{1}/{2}\;}}}{{{r}^{{3}/{2}\;}}+a{{M}^{{1}/{2}\;}}},
    \label{Omega angular velocity}
\end{equation}
where $u^\alpha=\left( {{u}^{t}},{{u}^{\phi }},{{u}^{r}},{{u}^{\theta }} \right)$ is the {four-velocity} of a particle.

% orbit_plot.png --------------------------------------------

\subsection{Radiative transfer}
\label{radiative transfer}
In this subsection we discuss the radiative transfer equation (RTE) and its general relativistic form (GRRTE). Thermal radiation is emitted from the hot AD and can have temperatures of $10^{7}\rm{K}$ or more, high enough to produce X-rays. This radiation spreads in the system and is absorbed by the disk material itself, material in possible outflow regions, escaping material propelled by winds, or even material stationary above the disk. The radiation specific intensity, ${{I}_{\nu }}\left( {\rm erg} \ {{\rm s}^{-1}} \ {{\rm cm}^{-2}} \ {{\rm ster}^{-1}} \ {{\rm Hz}^{-1}} \right)$, is thus adjusted and redefined depending on the trajectory it follows and the material it interacts with. Namely, the magnitude of the radiation increases as the trajectory passes through the hot and emitting disk material. At the same time, the intervening material absorbs part of the radiation, decreasing its intensity. Depending on this material's density and abundance, the light ray may manage to traverse the interceding material or else is absorbed along the way.

In classical environments \citep[see][]{RL}, the RTE for the specific intensity, $I_{\nu}$, at frequency $\nu \ \left(\rm Hz\right)$ takes the form
\begin{equation}
    {{I}_{\nu }}\left( s \right)={{I}_{\nu }}\left( {{s}_{0}} \right){{e}^{-{{\tau }_{\nu }}}}+\int_{{{s}_{0}}}^{s}{{{j}_{\nu }}\left( {{s}\,'} \right) {{e}^{- \left[ {{\tau }_{\nu }}\left( s \right)-{{\tau }_{\nu }}\left( {{s}\,'} \right) \right]}} {\rm d}{s}\,'} ,
    \label{rte}
\end{equation}
where the optical depth, $\tau_{\nu}$ (dimensionless quantity), is given by
\begin{equation}
    {{\tau }_{\nu }}\left( s \right)=\int_{{{s}_{0}}}^{s}{{{a}_{\nu }}\left( {{s}\,'} \right) {\rm d}{s}\,'}
    \label{opt.depth}
\end{equation}
and the absorption coefficient ${{a}_{\nu }}\left( {{\rm cm}^{-1}} \right)$ by
\begin{equation}
    {{a}_{\nu}}=n {{\sigma }_{\nu}} ,
    \label{a_nu R}
\end{equation}
with ${{\sigma }_{\nu }}\left( {{\rm cm}^{2}} \right)$ the absorbing area cross section at frequency $\nu$ and $n \left( {{\rm cm}^{-3}} \right)$ the material number density. Lastly, we have the emission coefficient, ${{j}_{\nu }}\left( {\rm erg} \ {{\rm cm}^{-3}} {{\rm s}^{-1}} {{\rm ster}^{-1}} {{\rm Hz}^{-1}} \right)$, which is given by
\begin{equation}
    {j}_{\nu }=\frac{{\rm d}{{I}_{\nu }}}{{\rm d} s} .
\end{equation}
In the above equations, $s$ stands for the classically measured path length.

When we wish to study the problem in general relativistic environments, several alterations must be made to the above equations to cope with the physical phenomena that arise, namely length contraction. As we explained in more detail in Part I \citep[Sect. 2.5]{KpartI}, the relativistic form of Eq. (\ref{rte}), the GRRTE, is then
\begin{equation}
    {{\mathcal{I}}_{\nu }}\left( \lambda  \right)={{\mathcal{I}}_{\nu }}\left( {{\lambda }_{0}} \right){{e}^{-{{\tau }_{\nu }}\left( \lambda  \right)}}-\int_{{{\lambda }_{0}}}^{\lambda }{\frac{{{j}_{\nu }}^{\prime }\left( \xi  \right)}{\nu { }\,'{{ }^{3}}}{{e}^{-\left[ {{\tau }_{\nu }}\left( \lambda  \right)-{{\tau }_{\nu }}\left( \xi  \right) \right]}}}{{\left. {{k}_{\alpha }}{{u}^{\alpha }} \right|}_{\xi }}{\rm d}\xi ,
\end{equation}
where $\mathcal{I}_{\nu }=I_{\nu }/\nu^3$ is the Lorentz invariant specific intensity, $\lambda$ the photon affine parameter \citep{B72,Wilkins72}, ${k_{\alpha}=\left( {{k}_{t}},{{k}_{\phi }},{{k}_{r}},{{k}_{\theta }} \right)}$ the photon covariant {four-momentum}, and $u^\alpha=\left( {{u}^{t}},{{u}^{\phi }},{{u}^{r}},{{u}^{\theta }} \right)$ the contravariant {four-velocity} of the environment fluid particle. The optical depth is then given by
\begin{equation}
    {{\tau }_{\nu }}\left( \lambda  \right)=-\int_{{{\lambda }_{0}}}^{\lambda }{{{a}_{\nu }}^{\prime }\left( \zeta  \right) {{\left. {{k}_{\alpha }}{{u}^{\alpha }} \right|}_{\zeta }}{\rm d}\zeta } ,
\end{equation}
where all quantities with a prime, such as ${j_{\nu }}^{\prime }$ and ${a_{\nu }}^{\prime }$, are measured in the emitting matter's rest frame. From the above equations, one can see that the intensity of the radiation is entwined with the material number density.

In the case where the disk material is very dense and opaque, the radiation is only emitted by a skin layer of the AD. This renders the solution of the GRRTE pointless, if not impossible. This is because the material density, and hence the optical depth, has so steep a gradient that numerical integration is notably challenging. Namely, the steeper the density increment is, the finer the affine parameter partition should be. When this increase is abrupt, however, the estimated intensity (the lower limit) is far from its actual value. This results in the need of an ever-denser partition mesh that is computationally unattainable. Consequently, another solution must be found, one that includes and accounts for all the relativistic effects brought on by the {spacetime} grid distortion due to the central mass. A suitable solution was presented in \citet*{AEL90} and \citet{ML96} for central radiation sources. We also studied this phenomenon much later in \citet{KC2014} and \citet{KpartI}, considering radiation emission from extended noncentral sources, possibly rotating, in Schwarzschild and in Kerr {spacetime}s.

In these circumstances, the suitable equation to solve in order to derive the received frequency-integrated specific intensity, $I_{\rm rec} \left( {\rm erg} \ {{\rm s}^{-1}} \ {{\rm cm}^{-2}} \ {{\rm ster}^{-1}} \right)$, using the emitted frequency-integrated specific intensity, $I_{\rm em}$, is
\begin{equation}
    {{I}_{\rm rec}}={{\left( \frac{{{g}_{tt, {\rm em}}}}{{{g}_{tt, {\rm rec}}}} \right)}^{2}}{{\left( \frac{1+{{\omega }_{\rm rec}} {{{k}_{\phi }}}/{{{k}_{t}}}\;}{1+{{\omega }_{\rm em}} {{{k}_{\phi }}}/{{{k}_{t}}}\;} \right)}^{4}} \frac{1}{{{\gamma }^{4}}{{\left( 1-{{V}^{{\hat{\phi }}}} \cos \psi  \right)}^{4}}} {{I}_{\rm em}} ,
    \label{I_rec}
\end{equation}
where $g_{tt,{\rm em}}$ and $g_{tt,{\rm rec}}$ are the metric components (Eq. \ref{metric}) at the point of emission and reception, respectively, and ${V}^{{\hat{\phi }}}={u}^{{\hat{\phi }}}/{u}^{{\hat{t}}}$ the material {three-velocity}. The first factor of the above product is a result of the time dilation between the point of emission and the point of reception. The second factor is used to take the effects brought on by {frame dragging} into account. Here, $\omega_{\rm rec}$ and $\omega_{\rm em}$ are given by Eq. (\ref{omega spacetime}) for the reception and emission point, respectively. Additionally, $k_\phi$ and $k_t$ are components of the photon momentum and are conserved quantities \citep{B72}. Finally, the third factor of Eq. (\ref{I_rec}) introduces the effects of the Doppler shift due to the emitting surface's motion. In this, $V={V}^{{\hat{\phi }}}$ is the source circular orbit {three-velocity}, $\gamma ={{\left( 1-{{V}^{2}} \right)}^{-{1}/{2}\;}}$ the emitting material Lorentz factor, and $\psi $ the angle between the emitting matter velocity and the photon emission direction.

% 3d_tori.png --------------------------------------------
   \begin{figure}
   \centering
   \includegraphics[width=\hsize]{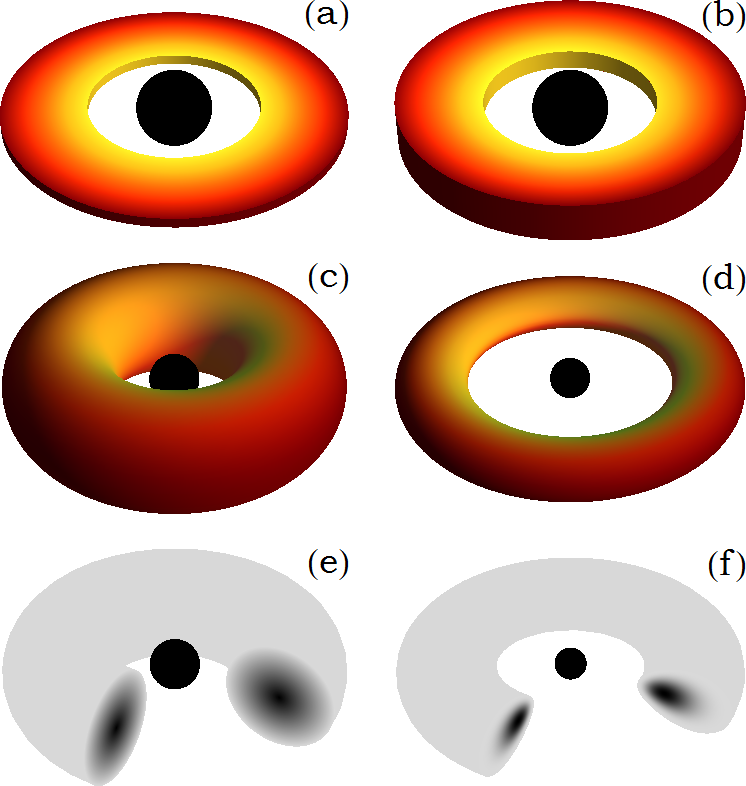}
      \caption{ADs we present results for. Disks (\textit{a}) - (\textit{d}) are optically thick, with temperatures higher for sections closer to the central BH {(Sect. \ref{accretion tori}a - d)}. Disks (\textit{e}) and (\textit{f}) are semi-opaque, with a higher density toward their material center {(Sect. \ref{accretion tori}e - f)}. The darker the material color in the cross sections appears, the higher the density is at that point. We can notice that the density increase in (\textit{e}) is smoother than in (\textit{f}).
      }
      \label{3d_tori}
   \end{figure}
   
\subsection{Accretion disks}
\label{accretion tori}
In this subsection we briefly review the AD models we considered for our calculations. Later on, we discuss and compare the results our computations revealed for these environments, their significance, and their implications.

The modeling of ADs is a large area of research, with many studies providing a broad spectrum of suggestions. For our work we consulted several recommendations for making the most effective choices. These include \citet{SS73}, \citet{Novikov-Thorne}, \citet{TP75}, \citet{A78,A88,A96}, \citet{K78}, \citet{NY94,NY95}, \citet{Esin97}, \citet{Beloborodov98,Beloborodov99,Beloborodov01}, \citet{Lasota99}, \citet{Igumenshchev}, \citet{NIA,NSPK}, \citet{Dubus}, \citet{FuerstPhD}, \citet{Noble07}, \citet{DGK}, \citet{NmC}, S{\k{a}}dowski (\citeyear{SadowskiApJ}, \citeyear{SadowskiPhD}), \citet{AF13}, and \citet{PSKN}.

We consider here six AD models, four of which are fully optically thick and two are semi-opaque. {The models presented are:}
{
\begin{enumerate}[label=(\alph*)]
    \item Slab (toy, snapshot model; opaque): a flat disk of half-height $h$, with range $r_{\rm ISCO} \leq r \leq 3r_{\rm ISCO}$. The cross section is a rectangle (Fig. \ref{3d_tori}a).\label{model_slab}
    \item Wedge (toy, snapshot model; opaque): a sloped disk of inner half-height $h$ with range $r_{\rm ISCO} \leq r \leq 3r_{\rm ISCO}$. The cross section is an isosceles trapezoid (Fig. \ref{3d_tori}b).\label{model_wedge}
    \item Torus (toy, snapshot model; opaque): a disk with a circular cross section with range $r_{\rm ISCO} \leq r \leq 3r_{\rm ISCO}$. The cross section center is at ${\left( r_c, \theta_c  \right)=\left( 2 {{r}_{\rm ISCO}}, {\pi }/{2} \right)}$ (Fig. \ref{3d_tori}c).\label{model_torus}
    \item Opaque rotationally supported torus (ORST; self-consistent model; opaque): a stationary and axisymmetric rotationally supported torus with its rotation vector collinear with the BH's (Fig. \ref{3d_tori}d, details in  \citealt{KpartI} Sect. 3.1f).\label{model_orst}
    \item Semi-opaque LFM torus (toy, snapshot model; quasi-opaque): A semi-opaque version of the torus model. The number density is maximized to ${{n}_{\rm c}}={{10}^{18}} {{\rm cm}^{-3}}$ at the cross section center (Fig. \ref{3d_tori}e). For this disk we have $a_{\rm abs} \cdot r_{\rm outer}\sim 1 - 5$, where $a_{\rm abs}$ the absorption coefficient and $r_{\rm outer}$ the disk's outer radius.\label{model_lfm}
    \item Semi-opaque radiation-pressure-supported polish doughnut (PD; self-consistent model; quasi-opaque): a stationary and axisymmetric disk \citep{A78,K78,YWF}. The maximum number density is ${{{n}_{\rm c}}={{10}^{18}} {{\rm cm}^{-3}}}$ at ${\left( r_c, \theta_c  \right)=\left( 12, {\pi }/{2} \right)}$. The density gradient increase is much more abrupt for PD than for LFM (we compare Figs. \ref{3d_tori}e and \ref{3d_tori}f). For this disk we also have $a_{\rm abs} \cdot r_{\rm outer}\sim 1 - 5$.\label{model_pd}
\end{enumerate}
\noindent The six models are displayed in Table \ref{Table0} and more information is available in the \href{https://gitlab.com/leelamichaels/aandrt-ii-online-material}{online material}.
}

% disk_velocities_II.png --------------------------------------------
   \begin{figure}[t]
   \centering
   \includegraphics[width=\hsize]{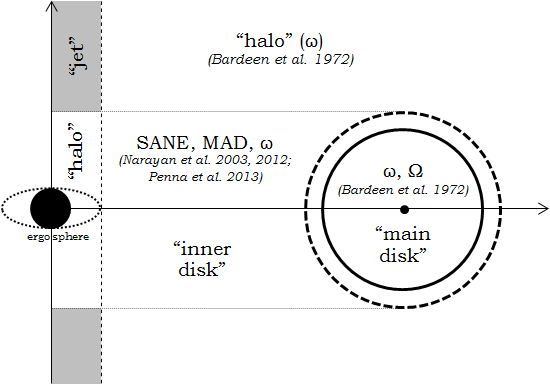}
      \caption{Miscellaneous velocity regions for the material we considered in the arrangement under examination. The dashed circle represents a thin region near the disk material edge, where matter is slowly escaping or accreting onto the main disk. %For this thin transition area, we additionally calculate the field, marked here with a target symbol; it moves with a velocity a negligible divergence away from that of the main disk's.
      {For this thin transition area, we additionally {calculated} the field recorded by a target particle moving with a velocity a negligible divergence away from that of the main disk’s.}      
        }
         \label{disk_velocities}
   \end{figure}

% Omega_Bundles_spins2.png --------------------------------------------
   \begin{figure*}[!th]
   \centering
   \includegraphics[width=\hsize]{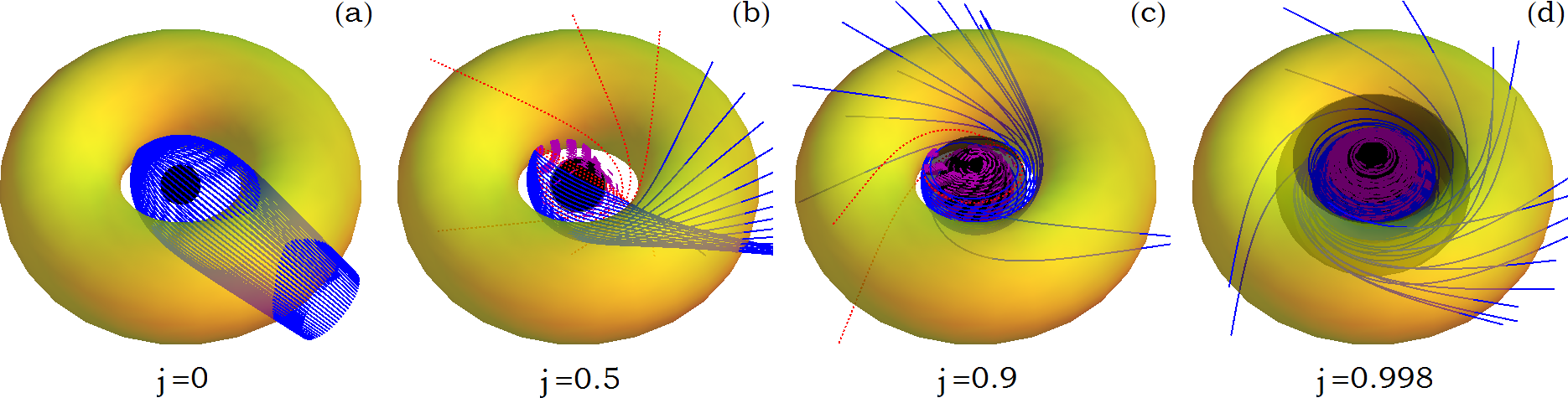}
      \caption{
      Output images of the photon bundle edition of \texttt{Omega} for four different values of the dimensionless angular momentum, $j$: 0, 0.5, 0.9, and 0.998, \textit{from left to right}. We see the BH event horizon (central black sphere), the ergosphere (gray ellipsoid or spindle torus), and the AD (yellow torus). Solid blue lines are trajectories of photons emitted from the AD. Dashed purple lines are trajectories that {would have to} originate from the event horizon, and dotted red lines are photons that arrive from outside the system. The latter two bring no additional radiation into the structure.
      }
      \label{Omega_Bundles_spins.png}
   \end{figure*}
 
\begin{table}
\centering \caption{Disk models considered.}

\begin{tabular}{cccc}
\hline \hline

 \text{Name} & \text{Type} & \text{Opt. thickness} & \text{Cross section} \\
 \hline
 \text{Slab} & \text{toy model} & \text{opaque} & \text{rectangle} \\
 \text{Wedge} & \text{toy model} & \text{opaque} & \text{trapezoid} \\
 \text{Torus} & \text{toy model} & \text{opaque} & \text{circle} \\
 \text{ORST} & \text{self-consistent} & \text{opaque} & \text{droplet-like} \\
 \text{LFM} & \text{toy model} & \text{quasi-opaque} & \text{circle} \\
 \text{PD} & \text{self-consistent} & \text{quasi-opaque} & \text{droplet-like} \\

\hline
\end{tabular}

\label{Table0}
\end{table}

For the material velocity, we looked into various studies and decided to split the configuration under examination into various regions with different velocity profiles. We show a schematic of the system sectors in Fig. \ref{disk_velocities}. “Halo” includes observers at rest in the local rest frame, rotating with $\omega$. “Main disk” is the stable, outer AD. The main disk matter can travel in {four} different ways: at rest in the local rest frame rotating with $\omega$, purely azimuthally with angular velocity $\Omega$ \citep{B72}, or by slowly infalling, such as in Standard And Normal Evolution (SANE), and Magnetically Arrested Disk (MAD) (see \citealt{NIA,NSPK,PSKN}). “Inner disk” includes the region of infalling material and thus contains only matter at rest in the local rest frame and infalling material that mimics SANE and MAD. The “jet” region is a section of a cylinder centered at the setup's rotation axis that includes the BH's ergosphere and thus has a radius $r=2M$. This region is only assumed to exist above a certain adjustable height, farther away from the BH. In that volume of space, we can have matter at rest in the local rest frame and matter flowing outward with a certain adjustable velocity. {Material in subregions} there may also be moving in different directions and at different speeds. For example, we can assume a narrower cone or cylinder with radius $r=M$ with a stronger outflow and a faster velocity, or even with a velocity of the opposite direction (see, e.g., \citealt{Asada16,Nakamura18,Park19}). 

\section{Results}
\label{Results}
In this section we show our results with pictures, density plots, and histograms. These include photon trajectories, radiation force magnitudes, AD images, BH spin estimations, and radiation-degenerated massive particle orbits. This material was created while studying assorted setups of disk models, spin parameters, velocity profiles, and more.

In each of the subsections we describe the results attained by the respective code and discuss their possible implications. Moreover, at the end of each part we briefly describe the potential improvements, extensions, and applications for the program in relative subjects.

\subsection{\texttt{Omega} results: Bundles of photons}
\label{Omega results}
\verb+Omega+ is the heart of most of our codes. It solves the differential equations of motion and studies massless and massive particle trajectories. Extensive work on particle orbits in BH environments and strong gravity can be found in, for example, {\citet{LPG}, \citet{LG}, and the following works}.

Since the latest version of \verb+Omega+ is not graphical, we show here the results of an earlier version designed and used to show photon bundle trajectories. In Fig. \ref{Omega_Bundles_spins.png} we see photon orbits reaching the same target-observer situated on the AD. Namely, moving backward, we trace the trajectory of a photon possibly received in a specific direction. If this path crosses the AD at any point, then it carries along radiation and energy. Under different circumstances, a photon would not reach the target carrying energy, since it would have to be emitted from the BH event horizon or from somewhere outside the system.

\begin{figure*}[!th]
   \centering
   \includegraphics[width=\hsize]{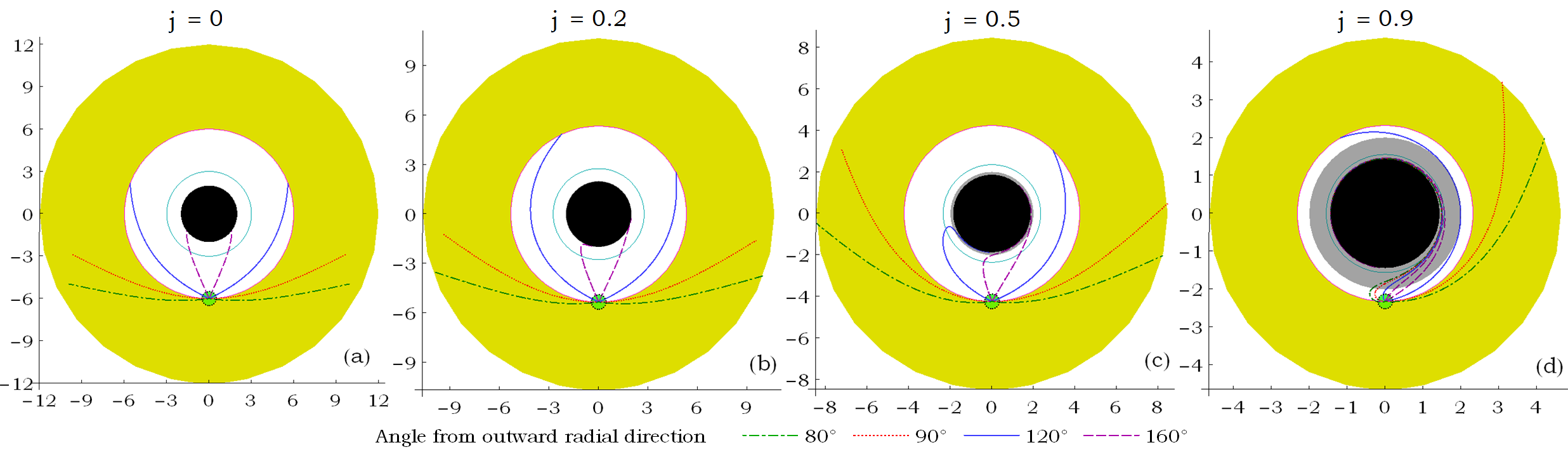}
      \caption{Equatorial photon trajectories launched from the AD material orbiting a clockwise-rotating BH with increasing dimensionless angular momentum, $j$ (\textit{from left to right}). We see the event horizon (central black disk), the ergosphere (gray annulus), the photon ring (inner cyan solid line circle), and the AD (outer thick yellow annulus) from $r_{\rm inner}=r_{\rm ISCO}$ to $r_{\rm outer}=2\:r_{\rm ISCO}$. {The photons {were} launched at angles measured from the outward radial direction at }
      %angles from different outward radial directions: 
      $80^{\circ}$ (dot-dashed green), $90^{\circ}$ (dotted red), $120^{\circ}$ (solid blue), and $160^{\circ}$ (dashed purple). %\LEt{ Verify that your intended meaning has not been changed.}
      We notice that as the BH spin increases, the photon trajectories that reach the observer at the bottom (green disk) get dragged along by the {spacetime} rotation.
      }
         \label{composite_orbits_spins.png}
   \end{figure*}

One of the study facets this code can be directly used for is the examination of particle trajectories and how they are affected by {spacetime} rotation. For example, in Fig. \ref{composite_orbits_spins.png} we see the effects of increasing BH spin on photons launched in different angles and moving in the equatorial plane. Moving to higher spins, we see that, as expected, the photons reaching the target move with the {spacetime} rotation itself (clockwise). The sky-scanning process and the {spacetime} rotation effects on photon trajectories can also be seen in {a video \href{https://www.youtube.com/watch?v=x195P-ShqCo}{here}.}

Regarding possible extensions for this program, numerous further studies can be made using \verb+Omega+. For example, it can be extended to study particle trajectories in X-ray binary systems, light rays in multiple-star systems, and more. Additionally, the capability of following photon trajectories in proximity to massive or compact stellar objects can be used to examine and analyze phenomena related to and influenced by gravitational lensing \citep{Refsdal64,BN92,NB96,Bozza2010}.

% composite_orbits_spins2.png --------------------------------------------

\subsection{\texttt{Infinity} results: Disk images and radiation forces}
\label{Infinity results}

\begin{table}
\centering \caption{\texttt{Infinity} code executions for the various disk models and spin parameters.}

\begin{tabular}{rrrrrrr}
\hline \hline
\text{BH}&\multicolumn{6}{c}{-----------------\ \ \ \ \ Disk model \ \ \ \ \ ------------------}\\

\text{spin} & \text{slab} & \text{wedge} & \text{torus} & \text{ORST} & \text{LFM} & \text{PD} \\
 \hline
0 & 168 & 170 & 163 & 184 & 496 & 454\\
0.5 & 182 & 177 & 176 & 188 & 478 & 484\\
0.9 & 156 & 158 & 148 & 179 & 424 & 488\\
0.998 & - & - & - & 191 & 414 & 488\\
 \hline
Sums & 506 & 505 & 487 & 742 & 1812 & 1914\\

\multicolumn{7}{c}{Total number of runs: 5966}\\
\hline
\end{tabular}

\tablefoot{The first four models are the opaque disks we described earlier. {The slab is discussed in Sect. \ref{accretion tori}a, the wedge in \ref{accretion tori}b, the torus in \ref{accretion tori}c, and the ORST in \ref{accretion tori}d. The remaining columns are the two semi-opaque disks. We can find LFM in \ref{accretion tori}e and PD in \ref{accretion tori}f}. Any of the results can be made available upon request.
}

\label{Table1}
\end{table}

\begin{figure*}
\includegraphics[width=\textwidth]{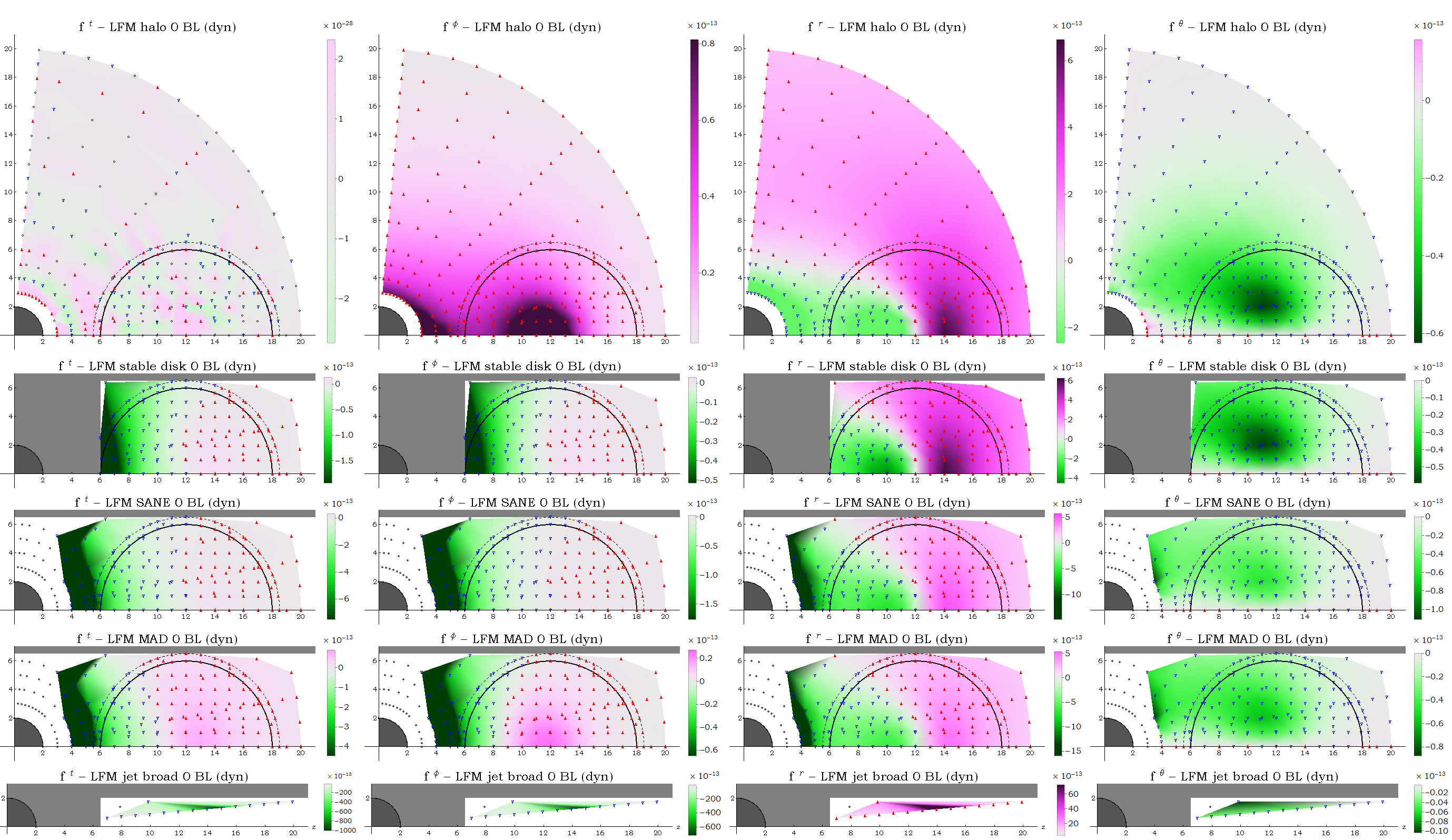}
\caption{{Radiation forces in $\mathrm{dyn}$ exerted on an electron} for the LFM semi-opaque disk model {(Sect. \ref{accretion tori}e)} for BH spin $a=0$ in BL coordinates. All axes use {geometrized} units, and the numbers shown are in $\mathrm{M}$. {The color schemes used, seen on the right of each plot, are centered at zero, namely a white diffuse color always {means that the radiation force is equal to zero}. We mark positive radiation forces with red filled "up" arrows and negative forces with blue or green empty "down" arrows. These marks act in the same manner as the diffuse colors and assist in the easier distinction of force signs in areas where radiation forces approach zero and hence plot colors approach white. Black empty circles denote a force exactly equal to zero (imperative for verification purposes) and gray dots denote points of null force values due to physical restrictions (e.g., the static limit or particular accreting material motion).} The \textit{first line} of plots (axes $x$-$z$) shows the radiation forces exerted on electrons at rest in the local rest frame (Eq. \ref{omega spacetime}). The \textit{second line} (axes $x$-$z$) shows forces on electrons moving with angular velocity $\Omega$ (Eq. \ref{Omega angular velocity}) in the appropriate {spacetime} region. The \textit{third} and \textit{fourth line} (axes $x$-$z$) show radiation forces exerted on electrons moving with velocity profiles that mimic the SANE and MAD models' material motion. The \textit{fifth line} (axes $z$-$x$) records radiation forces in an outflow region. The material there, a certain (adjustable) distance away from the horizon, additionally has a velocity component pushing it outward along the $z$-axis, as it would be moving in the outflow region of such systems. More spins and environments can be found in the \href{https://gitlab.com/leelamichaels/aandrt-ii-online-material}{online material}.}\label{LFMforce}
\end{figure*}

\begin{figure*}
\includegraphics[width=\textwidth]{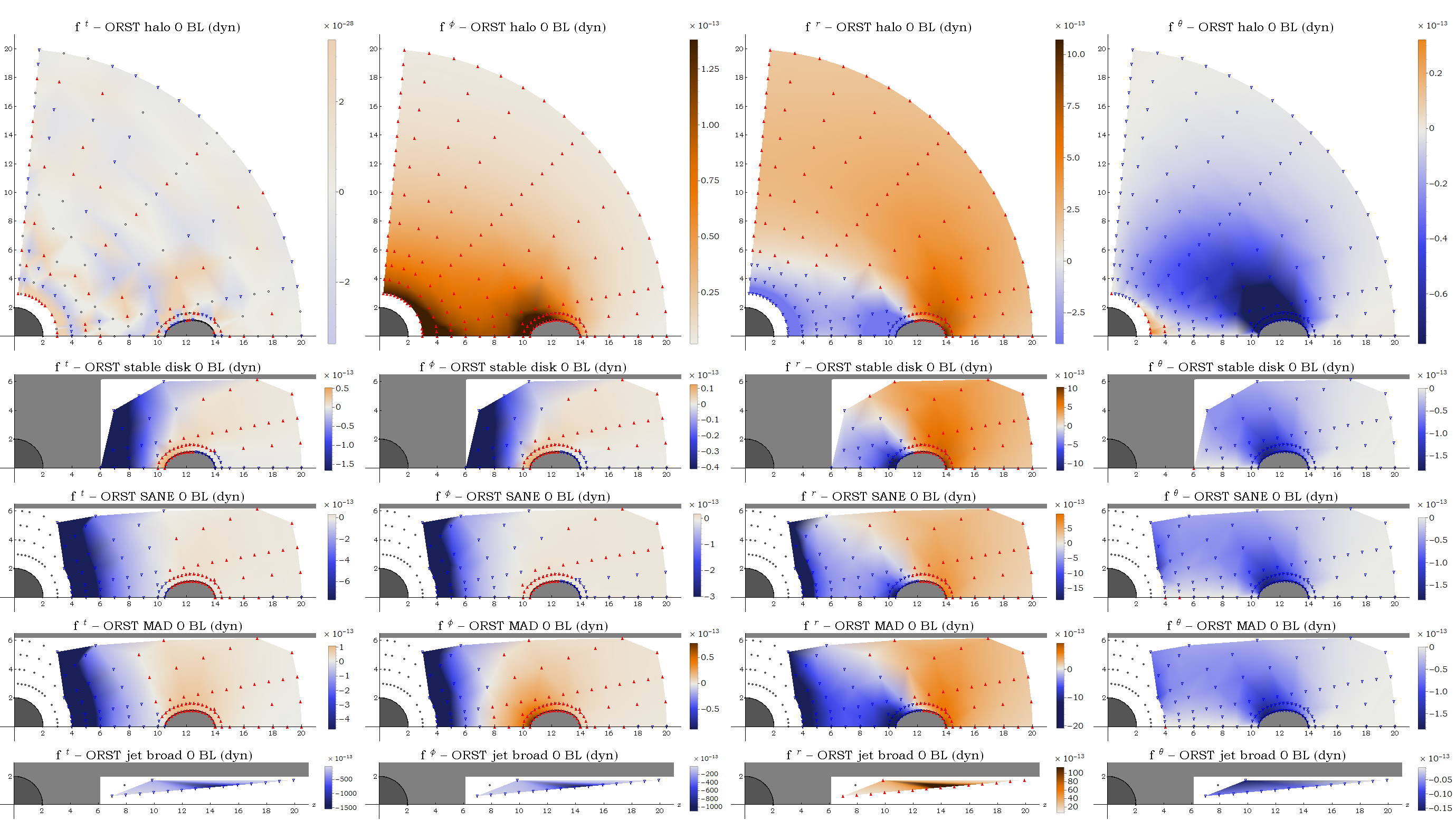}
\caption{{Radiation forces in $\mathrm{dyn}$ exerted on an electron} for the ORST disk model {(Sect. \ref{accretion tori}d)} for BH spin $a=0$ in BL coordinates. All axes use {geometrized} units, and the numbers shown are in $\mathrm{M}$. Plots are shown in the same manner as described for Fig. \ref{LFMforce}. More spins and environments can be found in the \href{https://gitlab.com/leelamichaels/aandrt-ii-online-material}{online material}.}\label{ORSTforce}
\end{figure*}

In order to study the radiation forces that appear at various points of the disk models described in Sect. \ref{accretion tori}, we developed the code \verb+Infinity+. Using this setup, we ran a large number of executions for all the models considered and with assorted BH spin parameters. The number of executions for each of these environments are listed in Table \ref{Table1}. These executions are for various target locations in each of the systems. They can be in the configuration's interior ($r<r_{inner}$) or exterior sectors ($r>r_{outer}$), on the surface of the AD or away from it, or inside the system outflow regions or scattered in the halo. Observers are thus spread throughout the entire volume of the configurations and record the incoming radiation and the respective force. Each of these execution targets can be seen {as dots and triangles} in Figs. \ref{LFMforce} and \ref{ORSTforce} and Figs. 2-7 of the online material. %\LEt{ Line colors and symbol types should be described in the figure legend and not repeated in the main text.}
More information on the selected AD characteristics can also be found in the \href{https://gitlab.com/leelamichaels/aandrt-ii-online-material}{online material}.

\begin{figure*}[t]
   \centering
   \includegraphics[width=\hsize]{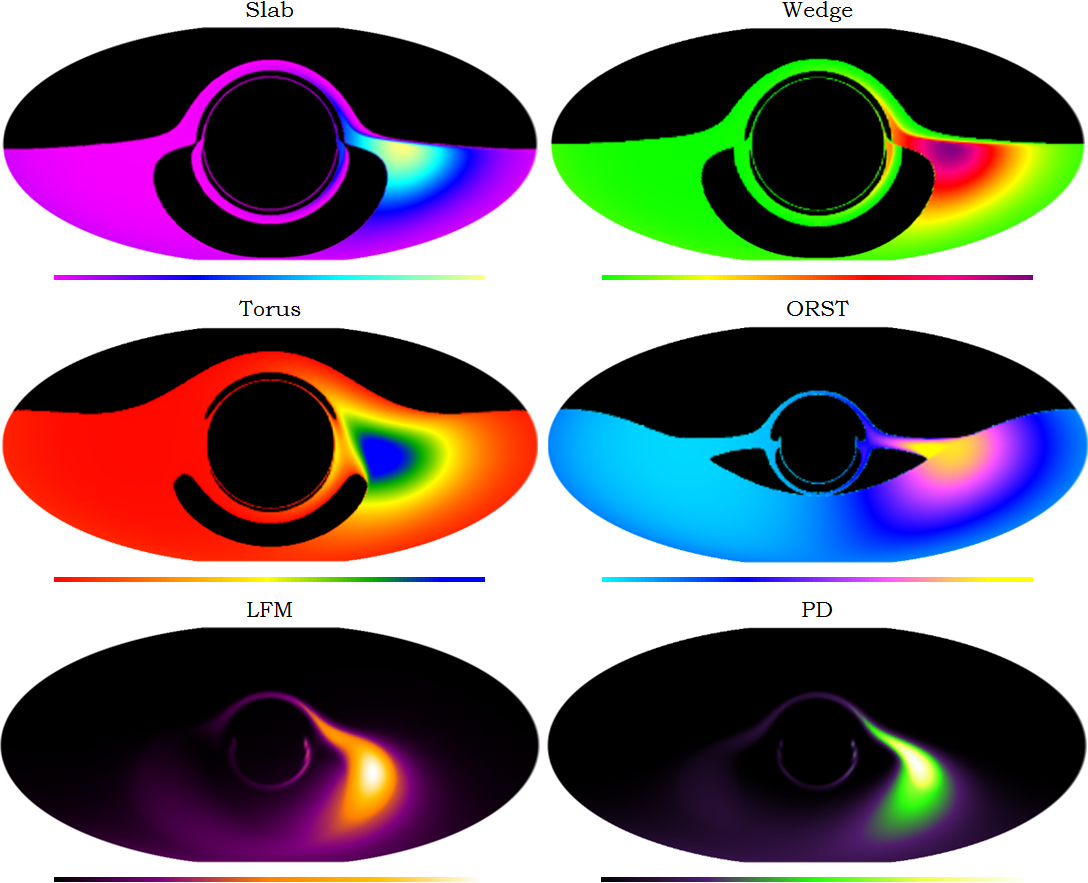}
      \caption{Mollweide projection sky maps of the frequency-integrated specific intensity, $I$. Each image shows the run results for a different disk model (Sect. \ref{accretion tori}a-f). Below each image, the model's color scale is displayed from minimum (\textit{left}) to maximum (\textit{right}). The circle of radiation in the center of each image is an {Einstein--Khvolson or "echo" ring} \citep{Sauer} created by photons traversing the disk area and reaching the observer after traveling above and below the BH horizon and AD. The radiation intensity difference between the left and right half of each image is due to the rotation of the AD. Radiation from material moving toward the observer is boosted, while radiation from material moving away is {deboosted}.
      }
         \label{mollweides3}
   \end{figure*}

In Fig. \ref{mollweides3} we see some Mollweide projection sky maps of the incoming radiation for the models mentioned in Table \ref{Table1}. In addition, videos of flights around, and if possible through, the disk for all the models we studied can be found on YouTube by searching for the name of this work's creator, {\href{https://www.youtube.com/playlist?list=PLilB2cSVqLeu6urpQ9u5Gk8LsA_O0UxRj}{"Leela Elpida Koutsantoniou"}}.

In Figs. \ref{LFMforce} and \ref{ORSTforce} we show result images for the radiation force magnitude and distribution for two of the ADs considered for spin $a=0$. In addition, in the online database, we show multiple large images with various values of $a$ that demonstrate the radiation four-force components for the previously mentioned ADs for various spin parameters. {The various disk models, and hence results, differ decisively from one another, and we thus use different colors for easier identification (i.e., for the model type and the force signs), but always denote zero radiation forces with a white color.}

{In these figures, we mark positive and negative radiation four-force components, distinguishing signs in areas where these forces approach zero. Furthermore, force values exactly equal to zero and null force values due to disk model or {spacetime} restrictions (e.g., static limit, accretion process rules) are also marked.} We finally note that each of these marks represents the results of one full code execution run. In some of the results depicted in these images, we applied an enhancement process \citep[see][Sect. 4.2]{KpartI} to increase the resolution by a factor of 25. We note, however, that it is also possible to change that to a higher or smaller number, or to remove it completely. It is additionally possible to run more than one application of the process.

Continuing on, we examine Figs. \ref{LFMforce} and \ref{ORSTforce} in detail. In Fig. \ref{LFMforce} we have a semi-opaque LFM disk and in Fig. \ref{ORSTforce} an opaque ORST disk. These density plots show the sign and magnitude of the BL four-force components in dyn ($\mathrm{1\,dyn=1\,g\cdot cm /s^{2}=10^{-5}\,N}$) at several points outside, near, and, if possible, inside some of the disk models mentioned in Table \ref{Table1} and Fig. \ref{mollweides3}. Each line of these pictures refers to target particles that belong to one of the aforementioned velocity groups: “halo” includes particles moving only with $\omega$ due to the {spacetime} rotation (Eq. \ref{omega spacetime}), “disk” includes particles in circular orbits with speed $\Omega $ (Eq. \ref{Omega angular velocity}), “SANE” and “MAD” refer to particles in inspiral orbits that mimic these two accretion situations, and “jet” refers to target particles that also have a radial outflow velocity component pointing outward.
   
% LFM0BLadd.png --------------------------------------------

% ORST0BL.png --------------------------------------------
   
% mollweides4.png --------------------------------------------

The first four lines of each picture are typical poloidal plots: the horizontal axis measures the cylindrical distance, $\varpi $, from the central axis, and the vertical measures the height, $z$, from the equatorial plane, which is also a symmetry plane for the ADs. The fifth line shows plots rotated 90 degrees: the horizontal axis is $z$ and measures the height from the equatorial plane, and the vertical axis is $\varpi $, measuring the cylindrical distance from the system rotation axis. The columns in each plot group show the distribution of the $t$-, $\phi $-, $r$-, and $\theta $- force components.

%The various disk model results differ decisively from one another, and we thus use different colors for easier identification {(i.e., for %negative force values, positive force values, and zero-force components). force signs}
 
Some of the results presented in Figs. \ref{LFMforce} and \ref{ORSTforce}, as well as the online database material, are expected, while others are not. Moreover, we note that the deviation of the BL halo ${{f}^{t}}$ plot from zero gives us an estimation of the numerical and computational errors present in our calculations. These estimated errors appear to be adequately below the forces we are attempting to calculate, so they are well within acceptable limits.

Looking at the first column of a picture group, we can see the effects of radiation on the energy absorption amount and rate from the target particles. The images provided by the simulations are in qualitative agreement with the expected results. For instance, we can see the energy transfer getting stronger the closer we get to the central object. In that region the absorbing and emitting particles of the material rotate and move faster in their trajectory. This means that the radiation is further beamed along the motion direction. In addition, since at these points we are outside the AD, the incoming radiation is getting further and further away from uniformity. These two aforementioned arguments explain why we expect the PR effect to be more potent there and to display its braking impact more strongly than in other areas. 

On the contrary, when examining areas farther outside the disk, the radiation is closer to uniformity. Thus, the PR effects are more easily concealed by the uniform and beamed radiation emitted by the moving material along its direction of motion. Additionally, we note that the farther away we move from the BH, the less curved {spacetime} is and, therefore, the less arced the photon trajectories get. We hence note the appearance of strong braking forces, $f^t < 0$, for areas with $\varpi \lesssim r_{\rm inner}$ and closer to the rotating $z$-axis, where $r_{\rm inner}$ is the inner edge of the disk. 
{As the cylindrical distance $\varpi$ from the rotation and symmetry $z$-axis increases beyond $r_{\rm inner}$, 
%As the cylindrical distance from the rotation increases and the symmetry axis, $\varpi$, becomes greater than $r_{\rm inner}$, 
the force turns positive, $f^t > 0$, but with a notably smaller magnitude.} %\LEt{ Verify that your intended meaning has not been changed.}
We also note that for semi-opaque ADs these observations continue to stand, but the zero-force surfaces move somewhat toward the configuration exterior. This is anticipated since adequate values for the optical depth are expected to accumulate for the photon trajectories %and 
{in order to} set in motion the processes described above.

\begin{figure}[t]
   \centering
   \includegraphics[width=\hsize]{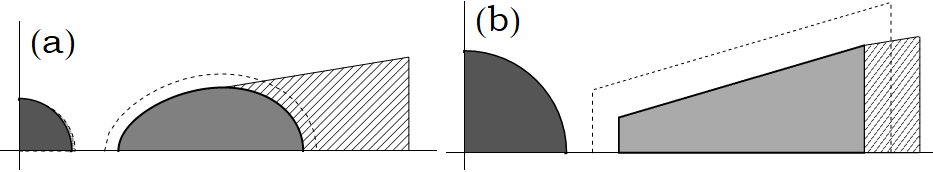}
      \caption{AD "shadows." Shown are partially shielded regions (in stripes) of the system next to opaque ADs. The disk opacity obscures emitting regions of the material, reducing the recorded radiation and thus unveiling the PR drag contribution.
      }
         \label{ADshadow}
   \end{figure}

The second column shows the azimuthal component of the radiation force. The {$\phi$-component} of the radiation force, $f^\phi$, follows the profile and distribution of the $f^t$ force in the vast majority of cases. This means that in the inner regions of the arrangements, within cylindrical radii $\varpi \lesssim r_{\rm inner}$, the force is negative, $f^\phi < 0$ (i.e., braking).%\LEt{ Verify that your intended meaning has not been changed.}
On the contrary, farther outward for $\varpi \gtrsim r_{\rm inner}$, the azimuthal radiation force turns positive, $f^\phi > 0$. Continuing farther out and departing from the setup, we record the expected decrease in radiation force magnitude and its eventual extinction. For opaque disk models we also note the presence of an additional area (Fig. \ref{ADshadow}). This area can perhaps be described as a "shadow" of the AD. Namely, for observers in regions outside the AD and at small heights from the equatorial plane, the azimuthal radiation force turns negative again. This happens because these areas are partially shielded from the bulk of the radiation emitted from the inner and hotter regions due to the opaqueness of the intervening material. This leads to the total received radiation magnitude being notably reduced. This alteration, in conjunction with the strong anisotropy of the observed radiation source, makes the PR effects more prominent and eventually dominant. We hence note the existence of a region that again displays braking forces in the direction of the disk rotation.

% AD_shadows2.png --------------------------------------------

The third column shows the radial four-force component, $f^r$, recorded for each disk model; {this} is also in qualitative agreement with the expected results. When the target is farther out from the disk, it receives more radiation coming from its interior local sky hemisphere than from its exterior one and is thus pushed farther outward. For a target farther inward from the disk, the opposite occurs: the outer hemisphere receives more radiation than the inner one, and the particle is pushed toward the BH.

The fourth column shows the {$\theta $-force}, $f^\theta$, in the poloidal plane and its effects on the structure particles. We note that the poloidal angle, $\theta$, is measured the typical way: $\theta=0$ for points on the positive $z$-axis, $\theta=\pi/2$ for the equatorial ($x$-$y$) plane, and $\theta=\pi$ for the negative $z$-axis. We first notice that, as expected due to symmetry, this force disappears in the equatorial plane. Additionally, we find it becomes increasingly negligible the closer we get to the vertical rotation axis. In almost all cases, this force is negative, which means it pushes the material upward toward the rotation axis. In the very few cases where the force sign is positive, {we detect these points being in small heights or in the immediate vicinity of the ergosphere or the event horizon.} 
%these points are in small heights \LEt{ I'm not sure what "in small heights" means.\ Verify that it is correct.} and in very close proximity to the ergosphere or the event horizon. 
At these points, the targets can only see a small part of the AD in the sky around them. As a consequence, the received radiation at these points is governed by the Einstein--Khvolson "echo" ring {of} the opposite side of the disk, across the BH. The targets there are thus pushed toward the equatorial plane, in contrast to the majority of cases. When we examine the bottom half of these configurations (i.e., for $z<0$), we notice the expected opposite behavior. The {$\theta$-force} there is positive almost everywhere, pushing material toward the $Oz'$-axis. Apart from these points, there is also a very limited quantity of points that record $f^\theta<0$ governed by the Einstein--Khvolson echo ring.

We next investigated the radiation forces documented in the central outflow regions of the systems. In these regions, a directed outflow or a jet could be formed, moving outward and away from the compact object, the AD, and the highly curved {spacetime}. For target particles moving in these areas, we assume two velocity profiles or more. We hence studied the radiation repercussions on material moving in the underlining {spacetime} grid and possibly outward. Before reviewing the results, we note that the radiation source is the AD and is situated behind the target particles as they move outward. Keeping this fact in mind, the results are qualitatively expected. In the radial direction, the source of light is inward, behind the target particles, and is thus driving the target particles outward ($f^r > 0$). For the poloidal direction, the emitting material is located below the targets and is consequently pushing the material toward the rotation axis ($f^\theta < 0$). As for the other directions, $t$ and $\phi$, we again see that they are consistent with each other and appear negative ($f^t < 0, f^\phi < 0$). This is also expected if we consider the nature and setup of the forces. Namely, for the azimuthal direction of motion of the emitted radiation, the light source is highly anisotropic and the bulk of the radiation is detected in angles at or around 90 degrees. This entails and explains the increased prominence of the PR braking effects.

We should note here, however, that the investigation is perhaps significantly dissimilar if we examine a slightly different environment: a more realistic outflow region of a BH. That is to say, we will perhaps obtain different results if we look into forces on material in an outflow that is not empty but contains blobs of matter. The radiation force profiles there are shifted and altered due to the existence of hot, emitting, and moving material before, after, or adjacent to the receiving target. {Such an examination should be carried out attentively in a future study.}

The force density plots (Figs. \ref{LFMforce} and \ref{ORSTforce} and the respective \href{https://gitlab.com/leelamichaels/aandrt-ii-online-material}{online material}) are useful for many reasons. First of all, they show us the magnitude and the sign of the radiation forces created by each disk model setup. They also reveal the differences between the models considered and their diversity. Furthermore, they provide us with novel and valuable information about the geometry and structure of the radiation field created in such objects. We can therefore see, for example, in which portions of the AD the forces are stronger or smoother. We can additionally notice the rearrangement of zero-force surfaces (whiter shades of colors in the plots) for the assorted disk models and its displacement for increasing values of the BH spin parameter. We can also see which specific force components have effects that influence and modify the stability and the evolution of the disk. We can investigate, for instance, whether the radiation has an impact and if it can modify the rotation of the AD material itself.

% barplotBL0.png --------------------------------------------
\begin{figure*}
\includegraphics[width=\textwidth]{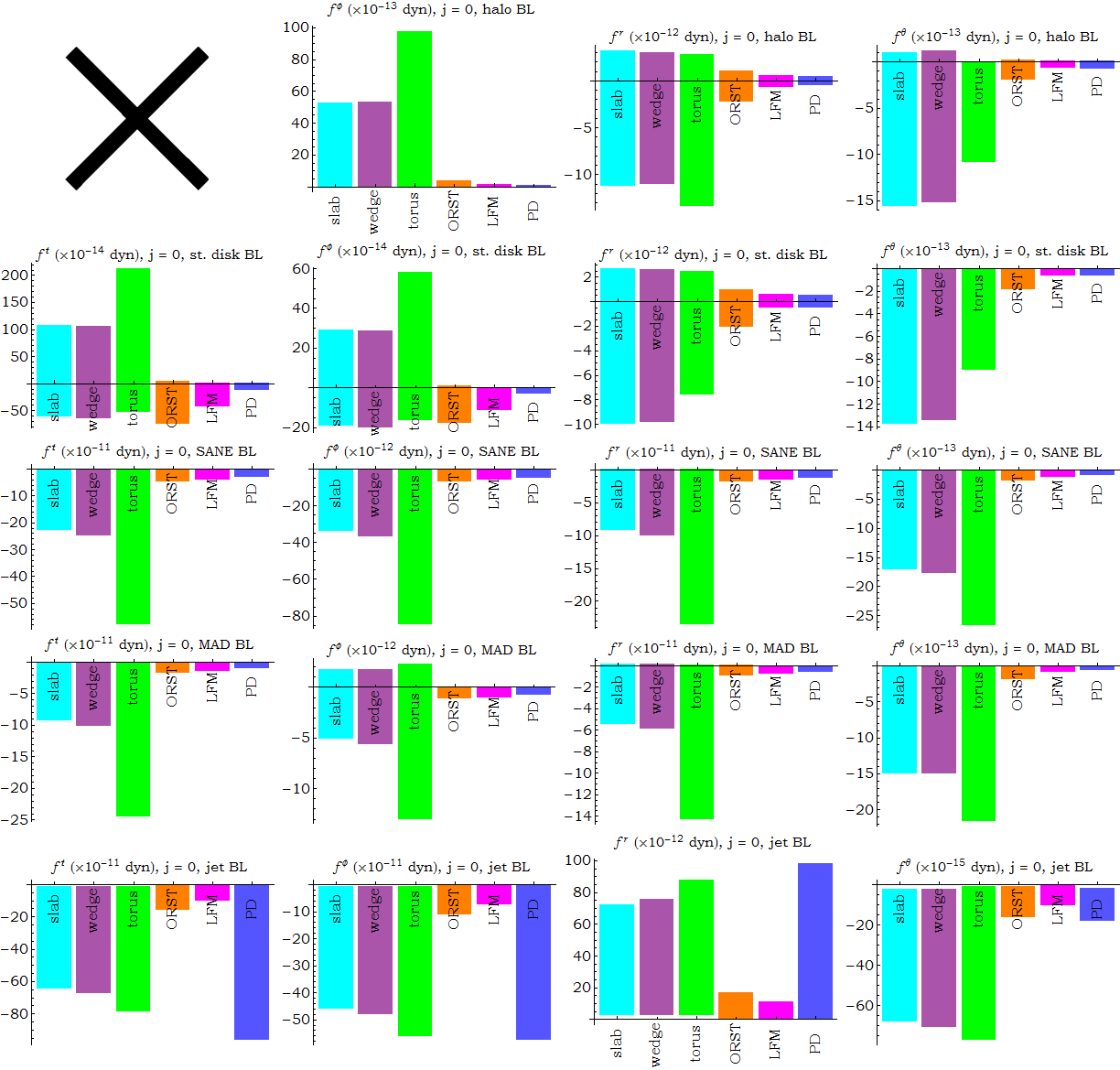}
\caption{Comparison of the radiation force components for the various AD models studied in $j=0$ {spacetime}. The models are described in Sect. \ref{accretion tori} and are shown in the plots in the same order. The halo $t$-force comparison (\textit{top left}) is not displayed since it only provides computational error information. More spins and environments can be found in the \href{https://gitlab.com/leelamichaels/aandrt-ii-online-material}{online material}.}
\label{barplot}
\end{figure*}

In addition to the aforementioned figures, we can examine the histograms that compare the radiation force components for the various models we studied. For example, we can see in Fig. \ref{barplot}, as well as the \href{https://gitlab.com/leelamichaels/aandrt-ii-online-material}{online material}, which setups record stronger forces.
We can hence examine the significance of the assorted disk characteristics and consider their results. We can in turn contemplate the importance of the disk volume versus its optical thickness, its vertical height, its rotation speed, its distance from the central compact object, and many other characteristics.

If we examine the force density plots and the comparative bar plots for all the disk models and spins considered, we can draw some conclusions about the recorded radiation. The Torus model, a voluminous and borderline opaque AD, appears to generate and exert the strongest radiative forces throughout its entire arrangement on all the disk or free material. Then follows the Wedge model, which is also opaque but with a much smaller vertical height and smaller radiative forces. Following closely, we have the opaque Slab model, which is similar to the Wedge model but with a constant vertical height throughout its structure. %These two models have very similar (and sometimes overlapping) characteristics and observed radiation effects. \LEt{ Verify that your intended meaning has not been changed.}
{These two models have similar characteristics and their observed radiation effect ranges often interweave. Namely, no model clearly surpasses the other one and their force component minima and maxima alternate between the two geometries.} We note, however, that overall Wedge records slightly higher radiation forces. We remark that this pair of models offers a unique opportunity to examine the radiation effects on the, perhaps puzzling, polar radiation forces and the diverse effects on material close to the rotation axis and the outflow region.

The model with the next-highest force magnitude is the opaque ORST, a model that significantly changes its size for increasing BH spin. Nevertheless, the disk's cross section center region only changes slightly and moves outward for increasing spin, $a$. Additionally, we observe in the bar plots that the geometrical characteristics of this model have in some cases interesting and unexpected consequences for the radiation forces exerted on the material (e.g., for $a=0.9M$).

After that, we have the semi-opaque disk models that demonstrate the production of softer radiation fields and follow the opaque disks in force magnitude. We first encounter the PD semi-opaque model. We notice that a setup of this kind appears to generate stronger forces in the outflow regions of the system compared to most of the other models. This is a subject well worth a more attentive look since it has important repercussions on the flow of mass and energy along the rotation axis. The effects of the radiation field should thus be reviewed in terms of their possible ramifications for and influence on the outflow and the jet, and more importantly on its collimation.

Along with the PD model, we have the semi-opaque LFM disk, which also creates a soft, but different, radiation field. This disk model has precisely the same geometrical attributes as Torus but is instead quasi-opaque. Comparing these two models, we can draw useful conclusions about the impact of the disk opacity and stratification. Namely, even though the two light sources are of the same size, the LFM disk exhibits a far weaker radiation field. We can thus experiment with the density gradient of the disk model and measure the consequences on the radiation field and the forces exerted. Moreover, by comparing the LFM disk with the PD, we notice again that even though LFM occupies a larger volume of space, the PD records larger radiation forces exerted on the material. These two points thus complement each other and agree with the aforesaid radiation notes. Most importantly, though, they hint that the primary factor responsible for the recorded forces is not the AD volume, but rather the material density increase rate.

Some other important notes for all the above figures are the following. %First of all, even though the forces may seem small or negligible, we should always keep in mind that these forces, ${\sim10^{-13}\mathrm{dyn}}$, {act upon the electrons and therefore a mass ${\sim10^{-27} \mathrm{g}}$}. 
{First of all, even though the forces may seem small or negligible, we should always keep in mind that they are exerted on the electrons and therefore a very small mass.
%This means that the (classical) acceleration due to radiation is ${\sim10^{14}\mathrm{cm/s^{2}}}$, and it is primarily acting upon the material electrons since ${f_{e}/f_{p}\sim(m_{p}/m_{e})^{2}}$. 
This means that the (classical) accelerations due to radiation are rather strong, and primarily acting on the material electrons since ${f_{e}/f_{p}\sim(m_{p}/m_{e})^{2}}$.}
This causes the material electrons to gradually brake, losing energy, or accelerate, gaining energy. This then results in electrons spiraling farther in toward the central object or being slung outward. The electrons' displacement then induces a charge separation between the material components. This then forces the protons (or ions) to follow the electrons into lower or larger orbits, subsequently causing parts of the AD material to slowly infall into the central BH or move outward. 

We remark here that the process of electrons changing rotation angular velocity and radius also leads to the generation of a ring electric current within the AD material. This current's central areas are situated near the disk's inner and outer edges. This hence results in a steady generation and accumulation of poloidal magnetic field loops in the setup. Depending on the material's physical properties, principally its magnetic Prandtl number, $Pr_m$\footnote{The magnetic Prandtl number, $Pr_m$ (dimensionless quantity), is the ratio of momentum diffusivity or viscosity to magnetic diffusivity.}, some parts of this material bring one polarity of these loops toward the center as they infall, while the opposite loop polarity is carried outward. Also taking into consideration the differential rotation of the AD, these magnetic field loops get twisted and then open up, filling the system with smaller magnetic loops and lines as described in \citealt{CKC06,manthos}. This magnetic field formation, {evolution, and possibly reconnection} is a very intriguing subject that is well worth an attentive investigation in a separate focused study.

Finally, looking closely at the density plots, one can notice that the $t$-force zero curves are not located at the same places as the $\phi$-, $r$-, and $\theta$-force zero curves. This was not an anticipated event, and it is worth mentioning explicitly since it gives rise to perhaps unexpected three-force components close to and through the disk.

In conclusion, there are many interesting possible extensions and usages of the \verb+Infinity+ code. We can further investigate the radiation and energy exchanges between various members of more complex arrangements. For instance, we can inspect the assorted transactions between partners of multiple-star systems, including main sequence stars, giants, {or} supergiants. These configurations may or may not contain ADs as well. The algorithm can also be used to take an attentive look into the electrical and consequently magnetic effects occurring in the systems. Our results revealed the generation of ring currents throughout the disk material due to the forces accelerating or braking the electron population. These developments thus lead to the generation of poloidal magnetic fields. The magnetic field lines in the structures later on extend, curl, twist, and open up, filling the volume of space inside the configurations, and possibly in areas farther out. These areas are thus filled with magnetic loops of the opposite curl. Keeping in mind the gradual extension of these opposite curl magnetic loops, further investigation can be conducted into the possibility and repercussions of magnetic reconnection. In addition, the particle trajectories and equilibria should be reconsidered, taking the new radiation and magnetic field environment conditions into account \citep*[see for example][]{BdFG}.
High resolution images of all the plots can be {found in the electronic form of the article and additional plots for various spins and disk models in the online material.}

% Elysium_runs.png --------------------------------------------
   \begin{figure*}
   \centering
   \includegraphics[width=\hsize]{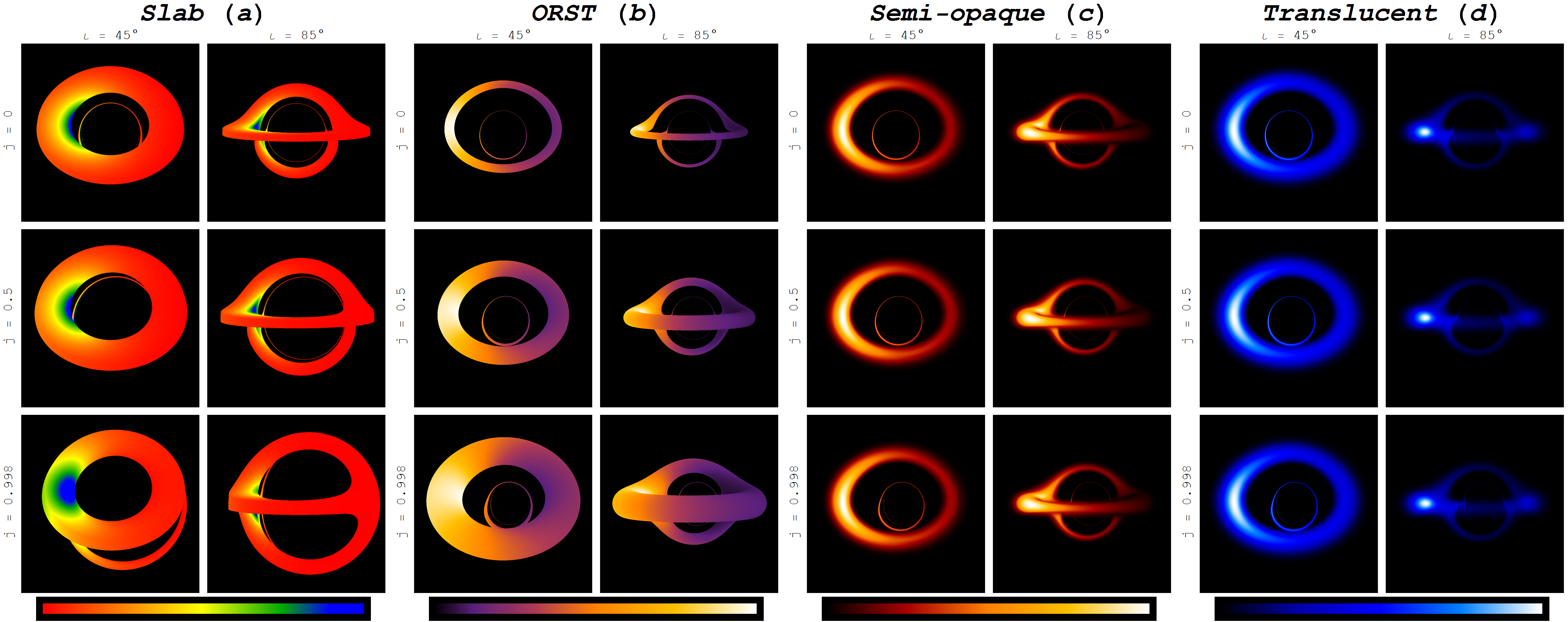}
      \caption{{Frequency-integrated specific intensity taken } %Photograph pictures 
      by running \texttt{Elysium} for (\textit{from left to right}) the opaque slab {(Sect. \ref{accretion tori}a), ORST (Sect. \ref{accretion tori}d), the semi-opaque {PD}, and the translucent {PD} (Sect. \ref{accretion tori}f)}. {On the \textit{left}} of each set of pictures, there are photographs taken from a screen with a $45{}^\circ $ inclination and {on the \textit{right}} with a $85{}^\circ $ inclination. Each line has a different BH angular momentum. Moving \textit{from top to bottom,} these are $j=0$, $j=0.5,$ and $j=0.998$. Below each picture group, we can see the used color scheme from minimum to maximum. The produced images are in agreement with other works, e.g., \citealt*{YWF}.
      }
         \label{Elysium_runs}
   \end{figure*}

\subsection{\texttt{Elysium} results: AD and BH photographs from infinity}
\label{Elysium results}

\verb+Elysium+ was created in order to construct BH and AD system images as they would be observed from farther away from the disk, or from "infinity." Similar work from different perspectives was also presented in, for instance, \citet{BLJun05}, \citet{Moscibrodzka14}, \citet{Davelaar18}, \citet{EHTa4}, and subsequent works. 

For our work, we ran several high resolution executions that show various AD models with different sizes, shapes, temperature profiles, and spins. Many other such models and environments could be studied using \verb+Elysium+, such as Novikov-Thorne ADs \citep{Novikov-Thorne}.
Here we show result images for different torus models of assorted spins and inclinations: in Fig. \ref{Elysium_runs} we see a slab disk in panel (\textit{a}), an ORST in panel (\textit{b}), a semi-opaque PD in panel (\textit{c}), and a very low absorption, translucent PD in panel (\textit{d}). 

\begin{figure}
   \centering
   \includegraphics[width=\hsize]{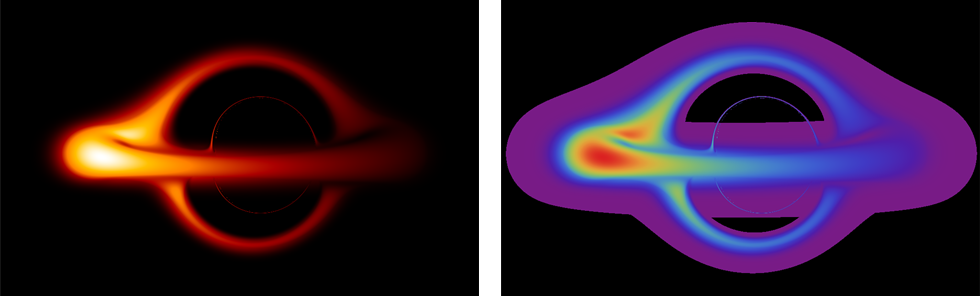}
      \caption{Generated picture of the same semi-opaque PD {(Sect. \ref{accretion tori}f)} shown in natural colors (\textit{left}) and in {pseudocolor} (\textit{right}). On the right, low density material is shown in purple instead of black, revealing the actual volume of the AD, which is significantly larger than what appears in natural colors on the left.
      }
         \label{invisible_torus}
   \end{figure}

We note here that black pixels in the aforementioned images do not necessarily signify the absence of material. This is due to the fact that, even in real observations, a very low density of material cannot be photographed. A typical example of such cases is the disk outer layers, as can be seen in Figs. \ref{Elysium_runs}\textit{c} and \ref{Elysium_runs}\textit{d}. To make this more apparent and to give a better idea of the actual AD volume, we show in Fig.\ref{invisible_torus} the photograph picture of the same AD in more natural colors (\textit{left}) and in {pseudocolor} (\textit{right}). In addition, we can notice {observationally expected} phenomena, such as limb darkening in Fig. \ref{Elysium_runs}\textit{c}.

\begin{figure}
\centering
\includegraphics[width=\hsize]{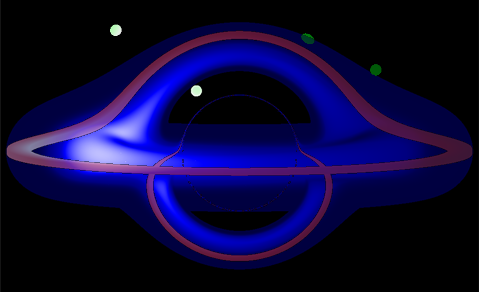}
\caption{Generated picture of an AD in multiple wavelengths {\citep[see e.g.,][]{UP95}}. The inner regions (electrons) are observed in X-rays ({blue}) and the outer, obscured, and colder regions (gas, dust, and molecules) in infrared ({red}). There {are also blobs of material in medium temperatures} above the disk, observable in optical wavelengths ({green}).
}
\label{multiwave}
\end{figure}

The application of this program is useful because, apart from simpler figures, it allows us to create the expected images produced by a multitude of observations. For example, we can have images created by combining high energy observations (e.g., X-rays) with lower energy observations (e.g., optical and radio; see, e.g., Fig.\ref{multiwave}). We can thus create images of objects with varying amounts of different components, such as dust grains, molecules, dense free electrons, {magnetic fields, and many more.}

Subsequently, we considered the feasible expansions of the \verb+Elysium+ code. In combination with the aforementioned extensions of the \verb+Omega+ routine, we can additionally enhance \verb+Elysium+ in order to carefully look into evidence of gravitational lensing. In addition to imaging disks around massive compact objects, we can add other systems, stars, or galaxies to the background, behind the compact object, and survey the gravitational lensing evidence and data. Another avenue of investigation is exemplified by Fig. \ref{Elysium_runs}. There, we can notice that the primary Einstein--Khvolson echo ring does not have a constant size, but rather depends on the distance of the disk from the event horizon. For instance, we notice the first echo ring of the leftmost disk to be larger than the rest and, in some cases, larger in size than the disk itself. This is because the inner edge of this disk model is further inward than in the other cases. This could be of use in certain cases, for example in cases where the main disk is too small, too close to the event horizon, or partially concealed by intervening material. In such cases, detection of the first Einstein--Khvolson echo can reveal evidence about the disk material and its distance from the central object despite the main disk not being visible or discernible.

   % invisible_torus2.png --------------------------------------------
   
% multiwave5.png --------------------------------------------

% Tranquillity_plot.png --------------------------------------------
   \begin{figure}
   \centering
   \includegraphics[width=\hsize]{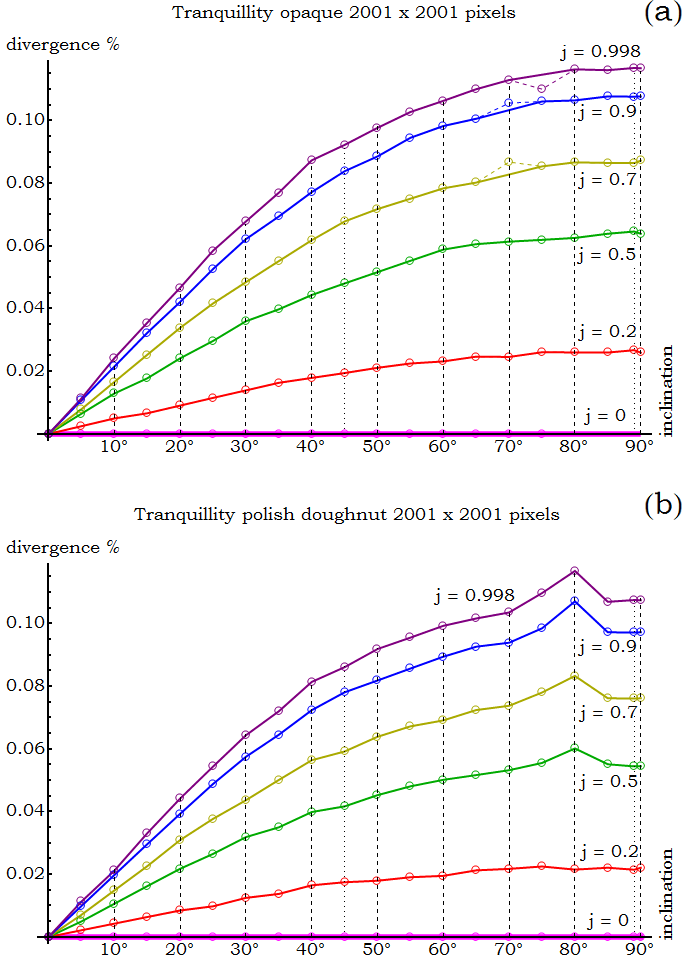}
      \caption{\texttt{Tranquillity} code collective results.\ The circles represent execution results. {The {ORST} (panel \textit{a}; Sect. \ref{accretion tori}d) changes significantly across various spins, while the {PD} (panel \textit{b}; Sect. \ref{accretion tori}f)} remains practically unchanged. There is a clear trend for the divergence evolution for every BH spin. In these plots, we also include the three worst code results, depicting them with the dashed colored lines.
      }
         \label{Tranquillity_plot}
   \end{figure}

\subsection{\texttt{Tranquillity} results: BH spin estimation}
\label{Tranquillity results}
\verb+Tranquillity+ is used to simulate and study images of disks rotating around a central object. The code examines images of arrangements at infinity and gives an estimation of the disk inclination relative to the line of sight and the central BH spin. Processing of a multitude of execution results leads then to the generation of divergence plots that can be used as a scale to promptly appraise the BH spin of a configuration.

In Fig. \ref{Tranquillity_plot} we can see two divergence plots, the results from the execution of 240 high resolution runs of \verb+Tranquillity+. The ADs under study are of varying inclinations, ranging from $0{}^\circ $ (face-on) up to $90{}^\circ $ (edge-on). In order to create these composite plots, the code was run for inclination angles every $5{}^\circ $, plus runs for an angle of $89{}^\circ $. We show here the study results of two very different tori. In the top plot, the AD changes significantly in size, inner and outer radius, and vertical height for each BH spin. On the contrary, in the bottom plot the torus size, shape, and location remain approximately the same for all spins. For both cases, there is a clearly visible trend for the divergence evolution of the same BH spin across the various inclinations.

\verb+Tranquillity+ shows good quality results for the inclination estimation \citep[see][Sect. 4.4]{KpartI}. We can thus use this attribute along with the aforementioned divergence plots to assess the central BH spin parameter. Results of such randomized tests can be seen in Table \ref{Table2}. There, we have a comparison between the actual AD inclination relative to the line of sight and the estimated disk inclination given by the code. Moreover, we have the comparison between the true BH spin parameter and the spin derived by the code. These two quantities appear to be in adequate proximity to each other, and the relative spin errors are limited and acceptable.

We considered the improvements that could be made to the \verb+Tranquillity+ code. We should remember that the results of this algorithm are affected by the AD model and its inner edge, but certainly do not rely on this convoluted and puzzling piece of information. In theory, this algorithm investigates how obtainable observational data can be used to provide an estimation of the central object spin, whatever that object might be. This means that the better and clearer the observational data are, the better the spin estimation is. The first step toward a better program is the creation of a larger information database. This will allow the code to automatically compare the observational input to a broad spectrum of statistics and discern the best fit and divergence scale plots for the observed disk. However, for such a complex and extensive undertaking, significant amounts of testing are required. In addition, cases with particular disk or {spacetime} geometrical situations should be recognizable and dealt with by the program automatically.

\begin{table}
\centering \caption{\texttt{Tranquillity} results and spin estimation.}
\begin{tabular}{lccccc}
 \hline
  \hline
 \text{Disk} & \text{Inclin.} & \text{Est. incl.} & \text{Spin} & \text{Est. spin} & \text{Spin error} \\
 \text{model} & \text{(${}^{\circ}$)} & \text{(${}^{\circ}$)} &  &  & \text{($\%$)} \\
  \hline
  
 \text{disk} & 15 & 16. & 0.13 & 0.13 & 2 \\
 \text{disk} & 72 & 72. & 0.49 & 0.49 & 0 \\
 \text{slab} & 9 & 9. & 0.06 & 0.07 & 8 \\
 \text{slab} & 79 & 79. & 0.25 & 0.24 & 3 \\
 \text{wedge} & 11 & 11. & 0.39 & 0.39 & 0 \\
 \text{wedge} & 86 & 86. & 0.4 & 0.39 & 3 \\
 \text{torus} & 39 & 39. & 0.12 & 0.14 & 13 \\
 \text{torus} & 64 & 64. & 0.32 & 0.32 & 0 \\
 \text{ORST} & 80 & 84.8 & 0.41 & 0.41 & 1 \\
 \text{ORST} & 82 & 86.3 & 0.86 & 0.81 & 7 \\
 \text{PD} & 15 & 14.9 & 0.98 & 0.98 & 0 \\
 \text{PD} & 48 & 47.5 & 0.75 & 0.76 & 1 \\

 \hline
\end{tabular}

\tablefoot{System properties that could be retrieved from observation-like images. We see the results for a sample of random parameter executions for all disk models. The AD inclination relative to the line of sight is adequately estimated by \texttt{Tranquillity}. Using this information in conjunction with the divergence plots created by the code, we have an estimation of the BH spin. The spin assessment results appear to be a satisfactory approximation of the true BH spin. 
}

\label{Table2}
\end{table}

\subsection{\texttt{Burning Arrow} results: Orbit degeneration due to radiation}
\label{Burning Arrow results}
   
% Burning_Arrow_composite2.png --------------------------------------------
\begin{sidewaysfigure*}[!htbp]
    \centering
    \includegraphics[width=\textwidth]{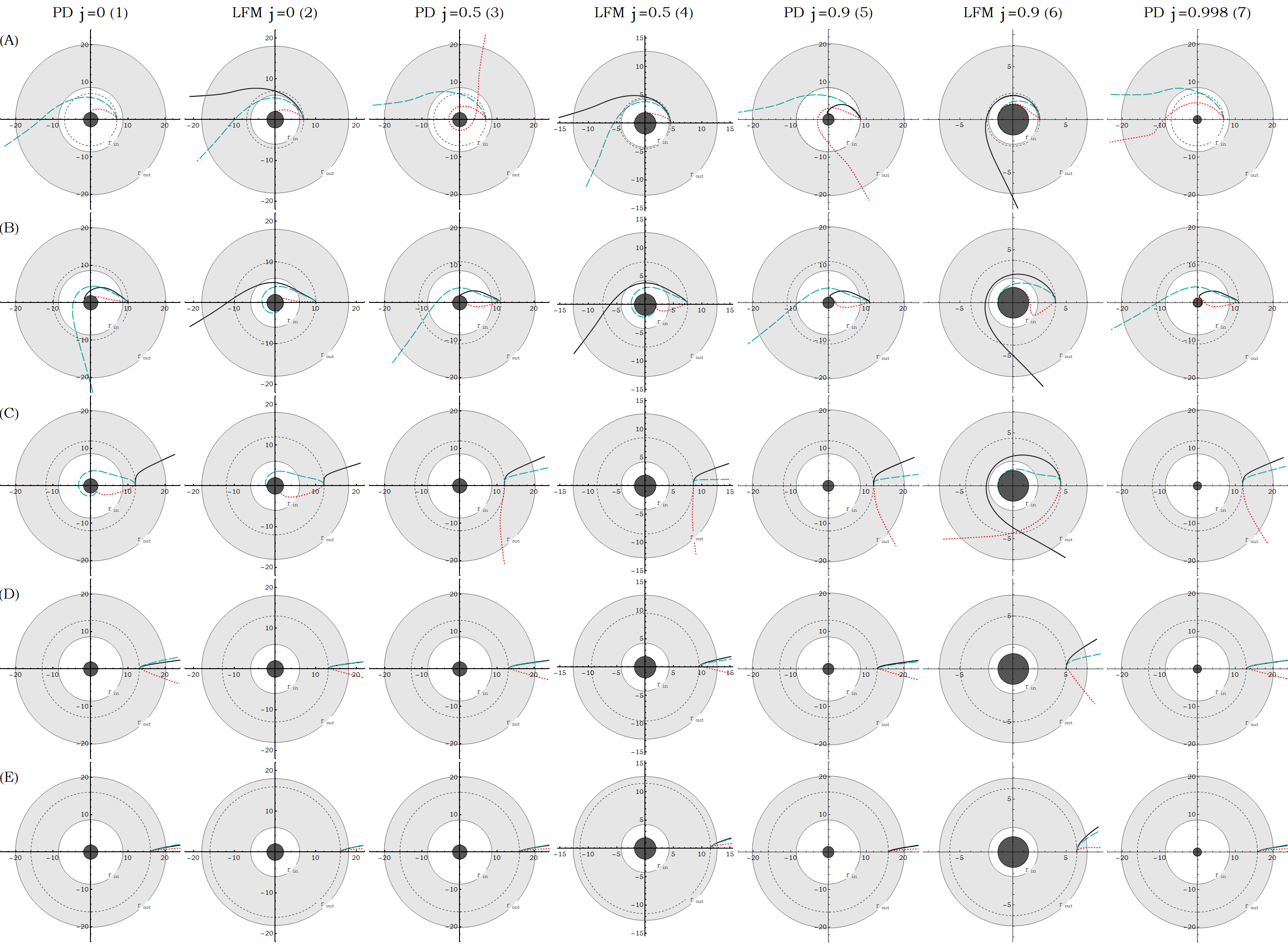}
    \caption{\texttt{Burning Arrow} results: Degradation of circular equatorial orbits due to the hot disk's radiation. We examine various BH spins and disk models and record the effects of radiation on electron trajectories. In solid black lines we have the target moving with the \citet{B72} velocity, in dashed cyan with the SANE velocity, and in dotted red with the MAD velocity.
    }
    \label{Burning_Arrow}
\end{sidewaysfigure*}

\verb+Burning Arrow+ was created to study the effects of the disk's radiation on the orbiting particles' motion itself. If we ignore radiation, matter in the AD rotates around the central object in a fashion similar to the circular orbits presented in \citet{B72} or relevant variations, such as SANE and MAD.

\verb+Burning Arrow+ reads and uses data from \verb+Infinity+ results regarding the radiation forces at various points in the system. It then calculates the acceleration components brought on by the disk's radiation. Finally, it determines and plots the consequent particle trajectory for assorted velocity profiles. This final particle trajectory now includes the additional radiation effects.

{We show in Fig. \ref{Burning_Arrow} some of our results for a {PD (Sect. \ref{accretion tori}f) and an LFM  (Sect. \ref{accretion tori}e)} disk at various spins.} We examine how electrons in equatorial trajectories in the AD around the BH are launched out of orbit due to the radiation forces. We remark that these emitted massive particles start from various radii from the axes' origin and thus each encounter different environments of material and radiation density.

We make note here of some of the "peculiar" and very interesting particle trajectories found in Fig. \ref{Burning_Arrow}. For example, we can see cases where the absorbing material is situated at the center of the local radiation field (e.g., Fig. \ref{Burning_Arrow}, panels \textit {C1}, \textit{C2}). We recall that, no matter the density or amount of radiation absorbed by a particle, the radiation flux, and hence the radiation force, is always a direct aftermath of the anisotropy of the radiation field \citep[see, e.g.,][]{RL}. Namely, no matter its intensity, if the radiation is isotropic, the radiation force is zero.

This has notable consequences, particularly when we examine particle motion in the local center of radiation. At such points, the radiation field is as close to isotropic as it can get for a moving target. This means that any deviation from the circular orbit (Eq. \ref{Omega angular velocity}), even a small one, results in the test particle getting into more and more anisotropic radiation regions. This makes orbits of this sort more and more unstable, since larger radiation forces are exerted. Trajectories with such an evolution thus get rapidly out of balance, and the moving particle is either slingshot outward or rapidly crosses the event horizon (e.g., when following velocity models such as SANE or MAD). We can thus say that the central area of radiation resembles an "unstable equilibrium point."

We also {record} cases where absorbing particles {have their trajectories significantly altered}. For example, there are cases where the particle trajectory starts from certain radiation force areas and then enters regions of stronger, softer, or counterbalanced forces. Other such notable cases are those where the traveling particle crosses into areas with markedly different {spacetime} curvature conditions. Instances of such situations can be seen, for example, in panels \textit{A5}, \textit{A7}, and \textit{C6} of Fig. \ref{Burning_Arrow}. These changes then lead to the appearance of visible and noteworthy instabilities in the trajectories, altering the expected particle curve.

Furthermore, we have cases where the particle trajectory approaches and possibly crosses the BH static limit, entering the ergosphere (e.g., Fig. \ref{Burning_Arrow}, panels \textit{B5}-\textit{B7}). This forces the particles under examination to change their rotation trend and follow the rotation of the central BH, as expected.

Finally, we examined possible extensions of the \verb+Burning Arrow+ code. Further applications to the particle trajectory adjustments due to the dynamical effects in the system should be considered and looked into. For example, as suggested by the force calculations, there are indications of the initial steps of charge separation. This is because of the braking or accelerating forces exerted on the material electrons. Since, however, a charge separation does not occur in the disk, we can examine the electrical effects taking place inside the material. Additionally, since we have the development of ring currents within the matter, further research should be conducted on the ensuing magnetic phenomena and how they could influence the material motion and equilibrium. Moreover, due to the aforesaid radiation forces altering the electron orbits and motion, a survey on the turbulence brought on by the radiation could be conducted. Ultimately, when examining the outer sections of the disk material, we notice that radiative accelerations appear stronger, while electrical binding appears weaker. Hence, in these areas, an electron orbit displacement is more likely to temporarily appear. In such a case, the development of instabilities and their ramifications for the magnetorotational instability and magnetic reconnection occurrence farther out should be {looked into thoroughly.}
   
\section{Conclusions and discussion}
\label{Conclusions}
In this work we have studied the radiation field produced by {the} hot AD orbiting a BH. This field proves to be eminently complicated for a number of reasons. {These reasons} include the existent general relativistic effects and the radiation source geometry, which is not located at the setup's center but extends into a large azimuthal and poloidal volume of space in the periphery. 

For this study we considered a broad variety of AD models. Some of the models we considered are opaque, while others are quasi-opaque with diverse degrees of density gradients. For the geometrical profiles of the disks, we examined an assortment of different shapes and sizes. We thus investigated situations of different disk characteristics, such as the maximum material height and the inner disk edge. Furthermore, we adopted and applied diverse velocity profiles for the material, attempting to cover as many of the particle motion situations as possible. Moreover, we examined different disk temperature profiles, such as disks heated up by the accretion process ($T\propto {{r}^{-{3}/{4}}}$) and isothermal disks. We note that the codes we created are highly adaptable, modifiable, and publicly available (links in Acknowledgements or \citealt{KpartI}). This means that adjustments and expansions can be made to accommodate any related research required for configurations in such {spacetime} conditions or in relevant arrangement setups.

A short summary of the research and results presented in this study is as follows. The \verb+Omega+ code and its multitude of generations provided results regarding massless and massive particle trajectories in the vicinity of compact objects. These objects affect and modify the {spacetime} form and structure around them in numerous ways. We can thus observe particle or bundle trajectories following the {spacetime} curvature and rotation. Subsequently, we used the \verb+Infinity+ code and thoroughly researched the radiation field created by the hot AD orbiting an astrophysical BH. Next we studied images of such setups from far away using the \verb+Elysium+ code. We scrupulously looked into the depiction results and attempted to locate the type of information we could draw for the configuration from each of the figures. This led to the development of the \verb+Tranquillity+ code. This algorithm collects arrangement depiction information and creates inclination-divergence plots. These plots {can later on be} used as a scale, providing an estimation of the spin of random environments of BHs encircled by ADs. So far, the results of the spin conjecture appear to be good approximations of the real spin of the compact objects. Finally, we investigated the repercussions of the disk's thermal radiation on the disk material itself. For this we used the \verb+Burning Arrow+ code and examined the destabilized trajectories followed by the disk electrons absorbing the high energy light.

{This work is novel in the sense that it is one of the very first times the radiation field created by the AD orbiting a BH is extensively studied, measured, and depicted.} This field was computationally calculated in a large volume of space for many configurations inward of, outward of and, if possible, inside the AD itself. Additionally, the radiation created was examined in the vicinity of notable system areas, including the disk's edges, the material outflow and inflow regions, the ergosphere, {the event horizon, etc.} A large quantity of numerical results, images, and plots have been produced and can be surveyed for many purposes.

The numerical results of this study show that optically thick accretion tori are more efficient at producing radiation forces for the material, as expected. Semi-opaque tori can, however, approach this limit under certain conditions. The results show that the primary factor affecting the recorded radiation forces appears to be the density gradient. Namely, considering disks with comparable structures, we see that the faster the material density increases, the larger the radiation forces exerted are. We clarify here that the quasi-opaque disks we used for this research are not tenuous. Their number density promptly rises to optical depths $\tau\geq 1,$ and thus we anticipate limited radiation losses due to a possibly smaller effective source extent. Furthermore, we observed opaque disk models of small volume recording higher radiation forces than significantly more voluminous quasi-opaque disks with optical depths $\tau \sim 1 - 5$. We thus consider the primary factor influencing the recorded radiation forces to be the material density increase rate and not the volume occupied by the disk material.

Also, we observe the anticipated stronger radiation forces when the torus maximum height increases. Comparisons between various disk models can additionally reveal the particular properties of each model. For example, we notice that for the outflow-jet regions, PDs and rotationally supported disks (ORST) record far larger radiation forces than the rest of the models. Specifically, for lower spins, $a$, where ORST is rather small scale, the PD prevails on the radiation forces exerted. On the contrary, for larger spins, ORST disks are more voluminous. This fact, in conjunction with the material full opacity, results in larger outflow forces being noted by ORST disks for larger spins. Additionally, concerning the ORST and PD models, we notice that for medium spins, $a \approx 0.5M$, rather large forces seemed to be exerted on halo free particles. Confirmation of this would require a closer investigation and, if confirmed, further inquiry on the reasons behind the phenomenon would be merited.

We also noticed the intense radiation force predominance of the ORST disk for very high spins. This is, however, overturned for LFM disk targets with the "main disk" velocity. This is anticipated since for that disk and velocity profile the material rotates faster and closer to the event horizon. When we examine softer material motion, such as SANE or MAD, the rotational velocities are nonetheless smaller and the ORST disk recaptures the leading position in radiation forces. A quick comparison between the considered opaque disk models reveals that the Torus model records the highest radiation force magnitude in the majority of cases. This is also anticipated since it is the most voluminous configuration of the opaque models. Apposing the above with a comparison between the large semi-opaque LFM disk and the smaller opaque disk results also upholds our hypothesis, that the primary factor deciding the radiation forces' impact on the disk material is the density gradient and not the disk volume.

Numerous parts of our work reveal that the effects of the thermal radiation on the disk material itself are significant and should not be disregarded. This radiation is larger and stronger than anticipated and creates {noticeable accelerations, positively not negligible.} Even though these radiation forces act upon the material electrons, they undoubtedly exert forces on the disk protons as well, even indirectly. Radiation thus results in material orbit destabilizations that affect the accretion process. These destabilizations could be in the form of turbulence, magnetorotational instability, or any other sort of instability. Depending on the destabilization area characteristics and properties, these processes can then lead to episodic events of matter accretion or expulsion. In a particularly interesting situation, the electron motion change leads to the formulation of electromagnetic effects, such as an electric ring current and an ensuing battery mechanism. Depending on the material's magnetic Prandtl number and magnetic diffusion, this battery mechanism can be of varying efficiencies. The cases where the magnetic field becomes strong enough to induce notable changes and further instabilities should thus be examined.

Another matter worth looking further into is the consequences of these significant radiation forces on the dynamics and the stability of the disk itself. The modifications brought about are noteworthy since long-considered concepts such as the ISCO must take radiation effect corrections into account.

An additional subject worth examining in the future is the radiation effects and possible repercussions for outflowing material close to the object's rotation axis. For instance, we could investigate the ramifications of radiation in the very early stages of material outward movement, which, according to our current results, seem appreciable. Additionally, the radiation density plots we show here demonstrate that the AD radiation promotes a tendency to {collimate the jet} material early on.

Finally, one more thing that could be considered in the future is how the calculated radiation affects more complex situations, such as the disk evolution in X-ray binaries and the observed hardness-intensity diagrams.

\begin{acknowledgements} 
The author would like to thank Stavros Dimitrakoudis, Dimitrios Millas, and George Pappas for their very useful questions and comments. The author would also like to express gratitude to the anonymous referees of Part I and Part II of this work for their valuable suggestions and guidance.
\newline
This work was in part supported by the General Secretariat for Research and Technology of Greece and the European Social Fund in the framework of Action “Excellence". Part of this work was performed at the Research Center for Astronomy and Applied Mathematics of the Academy of Athens.
\newline
The algorithms discussed in this work can be found here:
\newline
\verb+Omega+: \url{https://gitlab.com/leelamichaels/Omega.git}
\newline
\verb+Infinity+: \url{https://gitlab.com/leelamichaels/Infinity.git}
\newline
\verb+Elysium+: \url{https://gitlab.com/leelamichaels/Elysium.git}
\newline
\verb+Tranquillity+: \url{https://gitlab.com/leelamichaels/Tranquillity.git}
\newline
\verb+Burning Arrow+: \url{https://gitlab.com/leelamichaels/Burning\_Arrow.git}
\end{acknowledgements}

\bibliographystyle{aa}
\bibliography{AandRT2}

\end{document}